\documentclass[twocolumn]{aastex61}

\usepackage{bm}

\received{August 21, 2018}
\revised{\today}

\submitjournal{ApJ}

\shorttitle{BISTRO: The Magnetic Field of Perseus B1}
\shortauthors{Coud\'e et al.}

\begin{document}

\title{The JCMT BISTRO Survey: The Magnetic Field of the Barnard 1 Star-Forming Region}

\correspondingauthor{Simon Coud\'e}
\email{scoude@usra.edu}

\collaboration{The B-fields In STar-forming Regions Observations (BISTRO) collaboration}
\noaffiliation

\author[0000-0002-0859-0805]{Simon Coud\'e}
\affiliation{SOFIA Science Center, Universities Space Research Association, NASA Ames Research Center, M.S. N232-12, Moffett Field, CA 94035, USA}
\affiliation{Centre de Recherche en Astrophysique du Qu\'ebec (CRAQ), Universit\'e de Montr\'eal, D\'epartement de Physique, C.P. 6128 Succ. Centre-ville, Montr\'eal, QC, H3C 3J7, Canada}

\author[0000-0002-0794-3859]{Pierre Bastien}
\affiliation{Institut de Recherche sur les Exoplan\`etes (iREx), Universit\'e de Montr\'eal, D\'epartement de Physique, C.P. 6128 Succ. Centre-ville, Montr\'eal, QC, H3C 3J7, Canada}
\affiliation{Centre de Recherche en Astrophysique du Qu\'ebec (CRAQ), Universit\'e de Montr\'eal, D\'epartement de Physique, C.P. 6128 Succ. Centre-ville, Montr\'eal, QC, H3C 3J7, Canada}

\author[0000-0003-4420-8674]{Martin Houde}
\affiliation{Department of Physics and Astronomy, The University of Western Ontario, 1151 Richmond Street, London, ON, N6A 3K7, Canada}

\author[0000-0001-7474-6874]{Sarah Sadavoy}
\affiliation{Harvard-Smithsonian Center for Astrophysics, 60 Garden Street, Cambridge, MA, 02138, USA}

\author[0000-0001-7594-8128]{Rachel Friesen}
\affiliation{National Radio Astronomy Observatory, 520 Edgemont Rd., Charlottesville, VA, 22903, USA}

\author{James Di Francesco}
\affiliation{Department of Physics and Astronomy, University of Victoria, Victoria, BC, V8P 1A1, Canada}
\affiliation{NRC Herzberg Astronomy and Astrophysics, 5071 West Saanich Rd, Victoria, BC, V9E 2E7, Canada}

\author[0000-0002-6773-459X]{Doug Johnstone}
\affiliation{Department of Physics and Astronomy, University of Victoria, Victoria, BC, V8P 1A1, Canada}
\affiliation{NRC Herzberg Astronomy and Astrophysics, 5071 West Saanich Rd, Victoria, BC, V9E 2E7, Canada}

\author[0000-0002-6956-0730]{Steve Mairs}
\affiliation{East Asian Observatory, 660 N. A`oh\={o}k\={u} Place, University Park, Hilo, HI 96720, USA}


\author{Tetsuo Hasegawa} 
\affiliation{National Astronomical Observatory of Japan, National Institutes of Natural Sciences, Osawa, Mitaka, Tokyo 181-8588, Japan}

\author[0000-0003-4022-4132]{Woojin Kwon}
\affiliation{Korea Astronomy and Space Science Institute, 776 Daedeokdae-ro, Yuseong-gu, Daejeon 34055, Republic of Korea}
\affiliation{Korea University of Science and Technology, 217 Gajang-ro, Yuseong-gu, Daejeon 34113, Republic of Korea}

% ...


\begin{abstract}

We present the POL-2 850~$\mu$m linear polarization map of the Barnard~1 clump in the Perseus molecular cloud complex from the B-fields In STar-forming Region Observations (BISTRO) survey at the James Clerk Maxwell Telescope. We find a trend of decreasing polarization fraction as a function of total intensity, which we link to depolarization effects towards higher density regions of the cloud. We then use the polarization data at 850~$\mu$m to infer the plane-of-sky orientation of the large-scale magnetic field in Barnard~1. This magnetic field runs North-South across most of the cloud, with the exception of B1-c where it turns more East-West. From the dispersion of polarization angles, we calculate a turbulence correlation length of $5.0 \pm 2.5$~arcsec ($1500$~au), and a turbulent-to-total magnetic energy ratio of $0.5 \pm 0.3$ inside the cloud. We combine this turbulent-to-total magnetic energy ratio with observations of NH$_3$ molecular lines from the Green Bank Ammonia Survey (GAS) to estimate the strength of the plane-of-sky component of the magnetic field through the Davis-Chandrasekhar-Fermi method. With a plane-of-sky amplitude of $120 \pm 60$~$\mu$G and a criticality criterion $\lambda_c = 3.0 \pm 1.5$, we find that Barnard~1 is a supercritical molecular cloud with a magnetic field nearly dominated by its turbulent component.

\end{abstract}




\section{Introduction}
\label{sec:intro}

Magnetic fields, which are ubiquitous within the Galaxy \citep[e.g.,][]{Ordog2017, Planck2015XIX}, influence greatly the stability of molecular clouds and their dense filamentary structures in which star formation occurs \citep[e.g.,][]{Andre2014, Andre2015}. Specifically, magneto-hydrodynamic simulations have shown that a combination of magnetism and turbulence is needed to slow the gravitational collapse of molecular clouds, and thus decrease the galactic star formation rate \citep[e.g.,][]{Padoan2014}. Measuring the amplitude of magnetic fields in dense interstellar environments is therefore crucial to our understanding of the physical processes leading to the formation of stars and their planets.

Interstellar magnetic fields are difficult to observe directly. Early studies hypothesized that polarization of background starlight through the interstellar medium was due to the alignment of irregularly-shaped dust grains with magnetic field lines \citep{Hiltner1949}. Subsequent observations of thermal dust emission in the far-infrared \citep{Cudlip1982} showed polarization orientations nearly orthogonal to measurements in the near-infrared, supporting the picture of elongated dust grains. Although magnetic fields are considered the most likely cause of dust alignment in interstellar environments, the grain alignment mechanisms themselves still remain a theoretical challenge \citep[e.g.,][and references therein]{Andersson2015}.

The Radiative Alignment Torque (RAT) theory of grain alignment is currently one of the most promising models to explain the polarization of starlight towards clouds and cores \citep{Lazarian2007_review}. In summary, this model predicts that asymmetric, non-spherical dust grains rotate due to radiative torques from their local radiation field and then align themselves with their long axis perpendicular to the ambient magnetic field \citep{Dolginov1976, Draine1997, Weingartner2003, Lazarian2007a}. The degree of this alignment, however, depends on the quantity of paramagnetic material in the dust \citep{Hoang2016}. Submillimeter polarization observations of optically thin thermal dust emission will therefore lie perpendicular to the plane-of-sky component of the field. 

The B-fields In STar-forming Region Observations (BISTRO) survey aims to study the role of magnetism for the formation of stars in the dense filamentary structures of giant molecular clouds \citep{Ward-Thompson2017}. This goal will be achieved by mapping the 850~$\mu$m linear polarization towards at least 16 fields (for a total of 224 hours) in nearby star-forming regions with the newly commissioned polarimeter POL-2 at the James Clerk Maxwell Telescope (JCMT). With the unprecedented single dish sensitivity of the Sub-millimetre Common-User Bolometer Array 2 (SCUBA-2) camera on which POL-2 is installed, the BISTRO survey will significantly expand on previously obtained polarization measurements at submillimeter and millimeter wavelengths \citep[e.g.,][]{Matthews2009, Dotson2010, Vaillancourt2012, Hull2014, Koch2014, Zhang2014}.

Several of the star-forming regions observed by BISTRO are part of the Gould Belt, a ring of active star-forming regions approximately $350$~pc-across that is centered roughly $200$~pc from the Sun \citep{Gould1879}. Here, we present the BISTRO observations of the Barnard 1 clump (hereafter Perseus B1, or B1) in the Perseus molecular cloud ($d \sim 295$~pc; \citealt{Ortiz2018}). B1 is known to host several prestellar and protostellar cores at different evolutionary stages \citep[e.g.,][]{Hirano1997, Hirano1999, Matthews2006, Pezzuto2012, Carney2016}. This cloud was also a target of both the JCMT and \textit{Herschel} Gould Belt surveys (from 70~$\mu$m to 850~$\mu$m), thus providing a characterization of its dust properties \citep{Sadavoy2013,Chen2016}.

This paper presents the BISTRO first-look analysis of the Perseus B1 star-forming region. In Section~ \ref{sec:observations}, we first describe the technical details of the polarization observations, and outline the spectroscopic data used in this work. In Section~\ref{sec:results}, we show the POL-2 850~$\mu$m linear polarization map of B1 and its inferred plane-of-sky magnetic field morphology. We also characterize the relationship between the polarization fraction and the total intensity, and we compare the POL-2 data with previous SCUPOL observations. In Section~\ref{sec:analysis}, we explain our methodology for measuring the magnetic field strength from the polarization data, and then present the results of this analysis. In Section~\ref{sec:discussion}, we discuss the significance of these results for the role of the magnetic field on star formation within Perseus B1. Finally, we summarize our findings in Section~\ref{sec:conclusion}. 

\section{Observations}
\label{sec:observations}

\subsection{Polarimetric Data}
\label{sub:scuba2}

The JCMT is a submillimeter observatory equipped with a 15~m dish that is located at an altitude of 4,092~m on top of Maunakea in Hawaii, USA. Its continuum instrument is SCUBA-2, a cryogenic $10,000$ pixel camera capable of simultaneous observing in the 450~$\mu$m and the 850~$\mu$m atmospheric windows \citep{Holland2013}. The SCUBA-2 beams can be approximated by a two-dimensional Gaussian with a full-width at half-maximum (FWHM) of $9.6$~arcsec at 450~$\mu$m and $14.6$~arcsec at 850~$\mu$m \citep{Dempsey2013}.

The POL-2 polarimeter consists of a rotating half-wave plate and a fixed polarizer placed in the optical path of the SCUBA-2 camera (\citealt{Bastien2011}; \citealt{Friberg2016}; P.~Bastien et al. in prep.). POL-2 is the follow-up instrument to the SCUBA polarimeter (SCUPOL), which had a similar basic design \citep{Greaves2003}. While SCUBA-2 always simultaneously observes at both 450~$\mu$m and 850~$\mu$m, only the 850~$\mu$m capabilities of POL-2 were commissioned at the time of writing. In brief, POL-2 observes by scanning the sky at a speed of 8~arcsec~s$^{-1}$ in a daisy-like pattern over a field that is roughly 11~arcmin in diameter. Since the half-wave plate is rotated at a rate of 2~Hz, this scanning rate ensures a full rotation of the half-wave plate for every measurement of a $4$~arcsec box position in the map. For this paper, the Flux Calibration Factor (FCF) of POL-2 at 850~$\mu$m is assumed to be 725~Jy~pW$^{-1}$~beam$^{-1}$ for each of the Stokes $I$, $Q$, and $U$ parameters(the Stokes parameters are defined in Section~\ref{sub:polarization}). This value was determined by multiplying the typical SCUBA-2 FCF of 537~Jy~pW$^{-1}$~beam$^{-1}$ \citep{Dempsey2013} with a transmission correction factor of 1.35 measured in the laboratory and confirmed empirically by the POL-2 commissioning team using observations of the planet Uranus \citep{Friberg2016}.

Perseus B1 was observed with POL-2 between 2016 September and 2017 March as part of the BISTRO large program at the JCMT (project~ID: M16AL004). These observations total $14$~hours (or 20 individual sets of $\sim 40$-minutes observations) of integration in Grade~2 weather (i.e., for a 225~GHz atmospheric opacity, $\tau_{225}$, between $0.05$ and $0.08$). A 20-minute SCUBA-2 scan of B1 without POL-2 in the beam was also obtained on 2016 September 8 to serve as a reference for pointing corrections during data reduction.

The data were reduced using the \textsc{starlink} \citep{Currie2014} procedure \textit{pol2map}~\citep{POL2_Cookbook}, which is adapted from the SCUBA-2 data reduction procedure \textit{makemap} \citep{Chapin2013}. In particular, this routine is used to reduce POL-2 time-series observations into Stokes $I$, $Q$, and $U$ maps. We follow the convention set by the International Astronomical Union (IAU) for the definition of Stokes parameters. The default pixel size of the maps produced by \textit{pol2map} is 4~arcsec. For the analysis presented in this paper, we have instead chosen a pixel size of 12~arcsec at the start of the data reduction process to improve the resulting signal-to-noise ratio (SNR) of the final Stokes $I$, $Q$, and $U$ maps. 

The data reduction process is divided into three steps to optimize the SNR in the resulting maps: (1) the procedure \textit{pol2map} is run a first time without applying any masks to obtain an initial Stokes $I$ intensity map directly from the POL-2 time-series observations; (2) this initial Stokes $I$ map is then used as the reference for the automatic masking process of \textit{pol2map}, which is run a second time on the time-series observations to produce the final Stokes $I$ map; and (3) the masks obtained in Step~2 are also applied during a third run of \textit{pol2map} to reduce the Stokes $Q$ and $U$ maps, which are automatically corrected for the instrumental polarization. The uncertainties in each pixel of the Stokes $I$, $Q$, and $U$ maps are taken directly from the variance maps provided by the \textit{pol2map} procedure. The role of masking in the reduction of SCUBA-2 data, and incidentally POL-2 data, is discussed at length by \citet{Mairs2015}.

The correction for instrumental polarization is a crucial step in the analysis of any polarization measurement. If the instrumental polarization is not properly taken into account, then it may lead to erroneous results. For this reason, the latest model (January~2018) for the instrumental polarization of the JCMT at 850~$\mu$m was extensively tested by the POL-2 commissioning team with observations of Uranus and Mars (\citealt{Friberg2016,Friberg2018}; P.~Bastien et al., in prep.). They found that the instrumental polarization can be accurately described using a two-components model combining the optics of the telescope and its protective wind blind. While the level of instrumental polarization is dependent on elevation, it is typically $\sim$~1.5 per cent of the measured total intensity \citep{Friberg2018}.

We also use 850~$\mu$m polarization data of Perseus~B1 from the SCUPOL Legacy Catalog. \citet{Matthews2009} built this legacy catalog by systematically re-reducing SCUPOL 850~$\mu$m observations towards 104 regions, including previously published observations of B1 \citep{Matthews2002}, to provide reference Stokes cubes of comparable quality for all the astronomical sources with at least a 2~sigma detection of polarization. For this paper, the SCUPOL Stokes $I$, $Q$, and $U$ cubes for B1 were downloaded from the legacy catalog's online archive hosted by the CADC. To match the POL-2 results, we resampled the SCUPOL polarization vectors onto a 12~arcsec pixel grid. 

\subsection{Spectroscopic Data}
\label{sub:spectro}

The JCMT is also equipped with the HARP/ACSIS high-resolution heterodyne spectrometer capable of observing molecular lines between 325~GHz and 375~GHz (or 922~$\mu$m to 799~$\mu$m). The Heterodyne Array Receiver Program (HARP) is a $4 \times 4$ detector array that can be used in combination with the Auto-Correlation Spectral Imaging System (ACSIS) to rapidly produce large-scale velocity maps of astronomical sources \citep{Buckle2009}. In this paper, we use the previously published $\sim$14~arcsec resolution integrated intensity map of the $^{12}$CO J=3-2 molecular line towards Perseus B1 (project~ID: S12AC01) \citep{Sadavoy2013}. This intensity map was integrated over a bandwidth of 1.0~GHz centered on the rest frequency of the $^{12}$CO J=3-2 line at 345.796~GHz. The noise added by integrating over such a large bandwidth has no effect on the results presented in this work since the $^{12}$CO J=3-2 data is used only to indicate the presence of outflows in Figure~\ref{fig:fig1_b1_polarization}.   

It is important to note that SCUBA-2, POL-2, and HARP are not sensitive to exactly the same spatial scales. This difference is due to a combination of the different scanning strategies for each instrument and their associated data reduction procedures \citep[e.g.,][]{Chapin2013}. Hence, this difference must be kept in mind when combining results from different instruments, such as correcting for molecular contamination using HARP or comparing source intensities between POL-2 and SCUBA-2. While this difference is not an issue for the results presented in this paper, it may need to be taken into account in future studies using BISTRO data (see Section~\ref{sub:contamination} for more details).

Finally, this project makes use of spectroscopic data from the Green Bank Ammonia Survey (GAS) \citep{Friesen2017}. GAS uses the K-Band Focal Plane Array (KFPA) and the VErsatile GBT Astronomical Spectrometer (VEGAS) at the Green Bank Telescope (GBT) to map ammonia lines, among others, in nearby star-forming regions. In this work, we specifically use measurements of the NH$_3$ (1,1) and (2,2) lines towards Perseus~B1 (GAS Consortium, in prep.). These observations of NH$_3$ molecular lines at $\sim 23.7$~GHz have a spatial resolution of 32~arcsec and a velocity resolution of $\sim 0.07$~km~s$^{-1}$.

\section{Results}
\label{sec:results}

\subsection{Polarization Properties}
\label{sub:polarization}

The polarization vectors are defined by the polarization fraction $P$ and the polarization angle $\Phi$ measured eastward from celestial North. These properties are determined directly from the Stokes $I$, $Q$, and $U$ parameters, which is the commonly accepted parametrization for partially polarized light. The Stokes~$I$ parameter is the total intensity of the incoming light, and the Stokes $Q$ and $U$ parameters are respectively defined as $ Q = I \, P \, \text{cos}\left( 2 \Phi \right)$ and $ U = I \, P \, \text{sin}\left( 2 \Phi \right)$.

When $Q$ and $U$ are near zero, these values will be dominated by the noise in our measurements. This noise contribution always leads to a positive bias in the calculation of the polarization fraction $P$ due to the quadratic nature of the polarized intensity $I_P = [Q^2+U^2]^{1/2}$ \citep[e.g.,][]{Wardle1974, Montier2015, Vidal2016}. The amplitude of this positive bias can be approximated from the uncertainty $\sigma_{I_P}$ given in Equation~\ref{eq:polarised_uncertainty}, which is used in Equation \ref{eq:fraction} to de-bias the polarization fraction~$P$ \citep[e.g.,][]{Naghizadeh_Clarke1993}.

The de-biased polarization fraction $P$ (in per cent) can therefore be written as:
\begin{equation}
P = \frac{100}{I} \; \sqrt[]{Q^2 + U^2 - \sigma_{I_P}^2 } =  \frac{100}{I} \, I_P \, ,
\label{eq:fraction}
\end{equation}
where we re-define $I_P$ as the de-biased polarized intensity with uncertainty $\sigma_{I_P}$. This uncertainty $\sigma_{I_P}$ is given by:
\begin{equation}
\sigma_{I_P} =  \, \left[ \frac{\left(Q \, \sigma_{Q} \right)^2 + \left(U \, \sigma_{U} \right)^2}{Q^2+U^2} \right]^{1/2} \, ,
\label{eq:polarised_uncertainty}
\end{equation}
where $\sigma_Q$ and $\sigma_U$ are the uncertainties on the Stokes $Q$ and $U$ parameters respectively. The uncertainty $\sigma_P$ of the polarization fraction $P$ is given by:
\begin{equation}
\sigma_{P} =  \, P \, \left[ \left( \frac{\sigma_{I_P}}{I_P} \right)^2 + \left( \frac{\sigma_{I}}{I} \right)^2 \right]^{1/2} \, ,
\label{eq:fraction_uncertainty}
\end{equation}
where $\sigma_I$ is the uncertainty on the Stokes~$I$ total intensity.

Finally, the expression for the polarization angle $\Phi$ is:
\begin{equation}
\Phi = \frac{1}{2} \, \arctan \left( \frac{U}{Q} \right)  \, ,
\label{eq:angle}
\end{equation}
where $\Phi$ is defined between 0 and $\pi$ (0$^\circ$ and 180$^\circ$) for convenience, and its related uncertainty $\sigma_\Phi$ is given by:
\begin{equation}
\sigma_\Phi = \frac{1}{2} \, \frac{\sqrt{\left(U \, \sigma_{Q} \right)^2 + \left(Q \, \sigma_{U} \right)^2}}{Q^2 + U^2} \, .
\label{eq:error_angle}
\end{equation}

\subsection{BISTRO First-Look at Perseus~B1}
\label{sub:pol2_perseusb1}

\begin{figure*}
\includegraphics[width=0.495\textwidth]{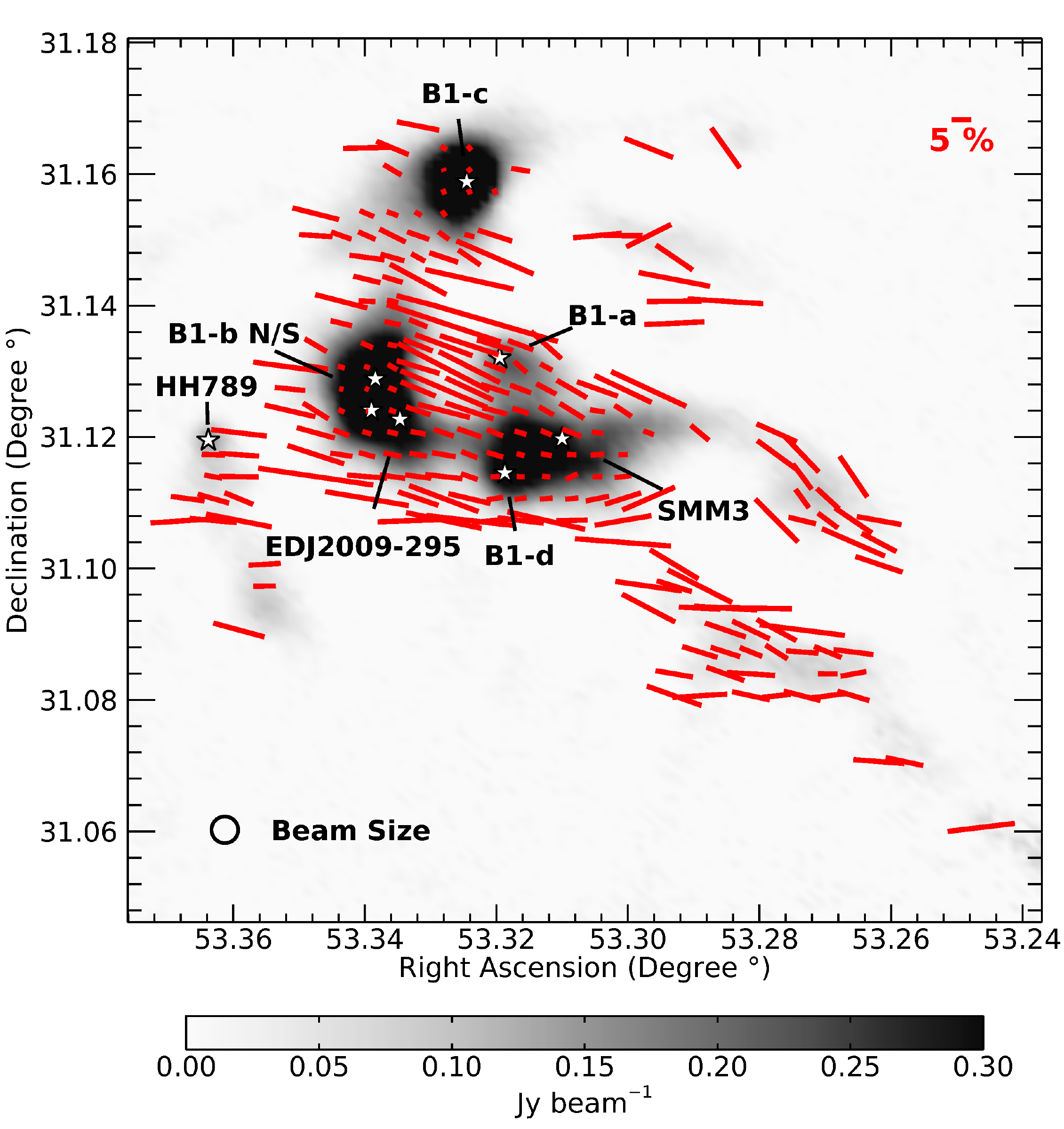}
\includegraphics[width=0.495\textwidth]{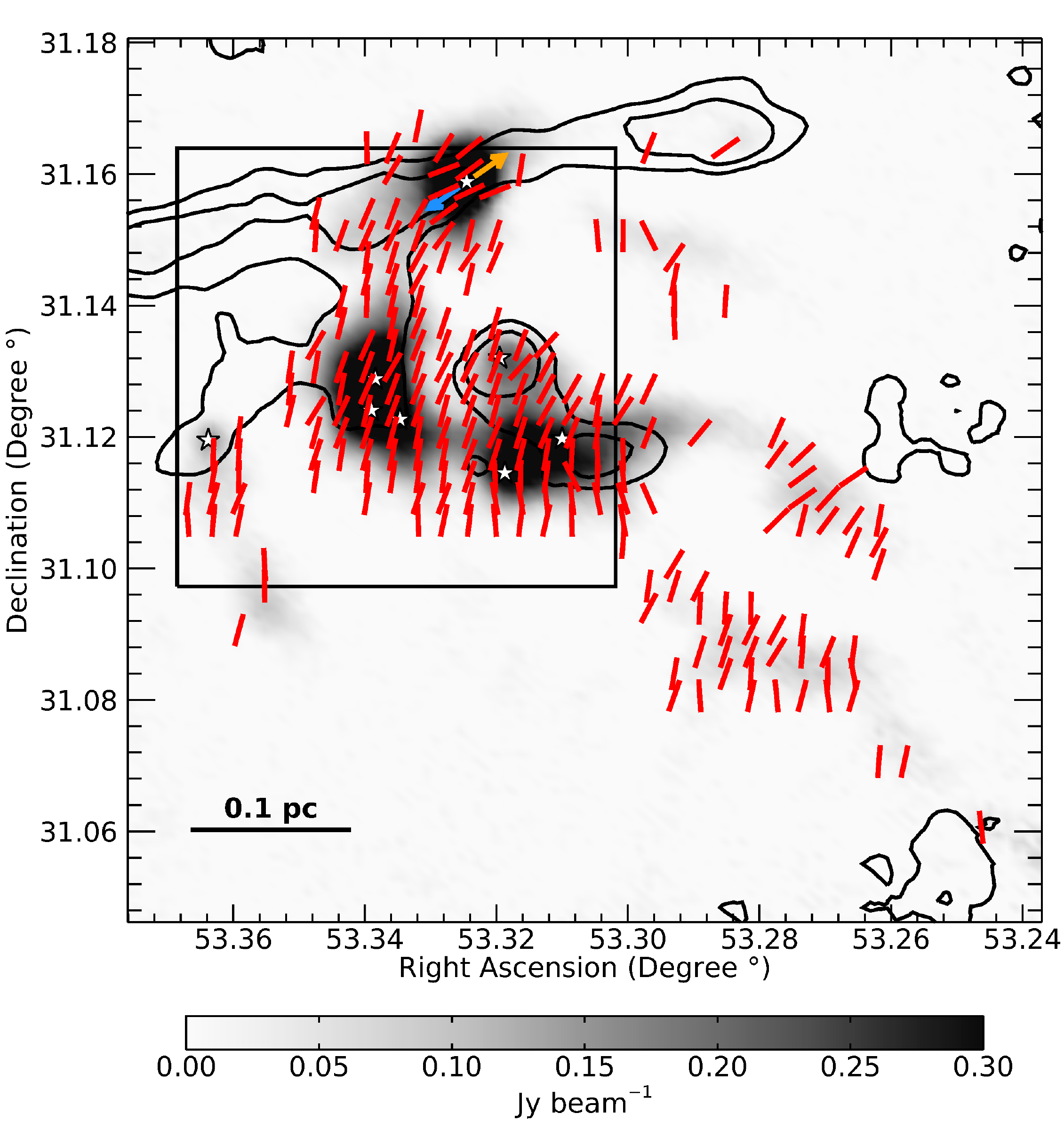}
\caption{The Perseus B1 star-forming region in 850~$\mu$m dust polarization from POL-2. In each panel, the gray scale indicates the measured Stokes $I$ total intensity. \textit{Left}: Vectors show the 850~$\mu$m linear polarization measured with POL-2 for a pixel scale of 12~arcsec, which is comparable to the effective beam size. The length of each vector is determined by its associated polarization fraction $P$ (per cent). The size of the SCUBA-2 beam at 850~$\mu$m (14.6~arcsec) is shown as a circle on the bottom left corner of the panel. Astronomical objects of interest are labeled and their positions are indicated by star symbols. \textit{Right}: Vectors show the inferred plane-of-sky magnetic field morphology obtained from the 90$^\circ$ rotation of the polarization vectors, which are normalized by length for clarity. The black contours trace the integrated intensity (10 K km s$^{-1}$ and 20 K km s$^{-1}$) of the $^{12}$CO~J=3-2 molecular line measured with HARP \citep{Sadavoy2013}. The blue and orange arrows around the protostellar core B1-c indicate the orientation of its blueshifted and redshifted outflows respectively, as characterized by \citet{Matthews2006}. Each lobe shows a clear bi-modal component with a FWHM of 5 to 10 km s$^{-1}$, and the typical velocity range in B1 is between -5 and 5 km s$^{-1}$ relative to the bulk of the cloud. The black box indicates the region analyzed for the improved Davis-Chandrasekhar-Fermi method described in Section~\ref{sub:dcf}. As a reference, the plain line drawn in the bottom left corner of the panel indicates a physical length of 0.1~pc.}
\label{fig:fig1_b1_polarization}
\end{figure*}

Figure~\ref{fig:fig1_b1_polarization} (left) shows the BISTRO 850~$\mu$m linear polarization map of Perseus B1 for a pixel size of 12~arcsec. The catalog of polarization vectors is calculated for every pixel of the POL-2 Stokes $I$, $Q$ and $U$ maps, but only vectors passing a set of pre-determined selection criteria are shown. These selection criteria are: a SNR of $I/\sigma_I\!>\!3$ for Stokes $I$ and its uncertainty $\sigma_I$, a SNR of $P/\sigma_P\!>\!3$ for the polarization fraction $P$ and its uncertainty $\sigma_P$, and an uncertainty $\sigma_P\!<\!5$~per cent for the polarization fraction. The criterion of $\sigma_P\!<\!5$~per cent was chosen arbitrarily as a precaution against potentially spurious vectors with anomalously high polarization fractions. These criteria provide a catalog of 224 polarization vectors for Perseus~B1.

The mean values of the Stokes uncertainties $\sigma_I$, $\sigma_Q$, and $\sigma_U$ for the polarization vectors shown in Figure~\ref{fig:fig1_b1_polarization} are 1.6~mJy~beam$^{-1}$, 1.3~mJy~beam$^{-1}$, and 1.3~mJy~beam$^{-1}$ respectively. At best, we achieve a sensitivity of $0.1$ per cent in polarization fraction and an uncertainty of $2.1^\circ$ in polarization angle, with mean values for $\sigma_P$ of $1.9$ per cent and for $\sigma_\Phi$ of $5.7^\circ$ for the entire catalog of vectors.

Assuming that interstellar dust grains are aligned with their long axis perpendicular to the magnetic field, the plane-of-sky field morphology in Perseus B1 is obtained by rotating the vectors in the polarization map by $90^{\circ}$. Figure~\ref{fig:fig1_b1_polarization} (right) shows the inferred plane-of-sky magnetic field map for B1. To help highlight the magnetic field structure, the rotated vectors are normalized to the same length. A contour plot of the HARP $^{12}$CO J=3-2 integrated intensity map from the JCMT Gould Belt Survey \citep{Sadavoy2013} is also included in the right panel of Figure~\ref{fig:fig1_b1_polarization}.

Selected submillimeter sources are identified in both panels of Figure~\ref{fig:fig1_b1_polarization} to serve as references for the discussion in Section~\ref{sec:discussion} \citep{Bally2008}. These sources are embedded young stellar objects which have been associated with molecular outflows \citep{Hatchell2009, Evans2009, Hirano2014, Carney2016}. Specifically, the lobes of the precessing molecular outflow originating from the protostellar core B1-c \citep{Matthews2006} are particularly well defined by the $^{12}$CO J=3-2 contour plot shown in the right panel of Figure~\ref{fig:fig1_b1_polarization}. 

\begin{figure}
\includegraphics[width=\columnwidth]{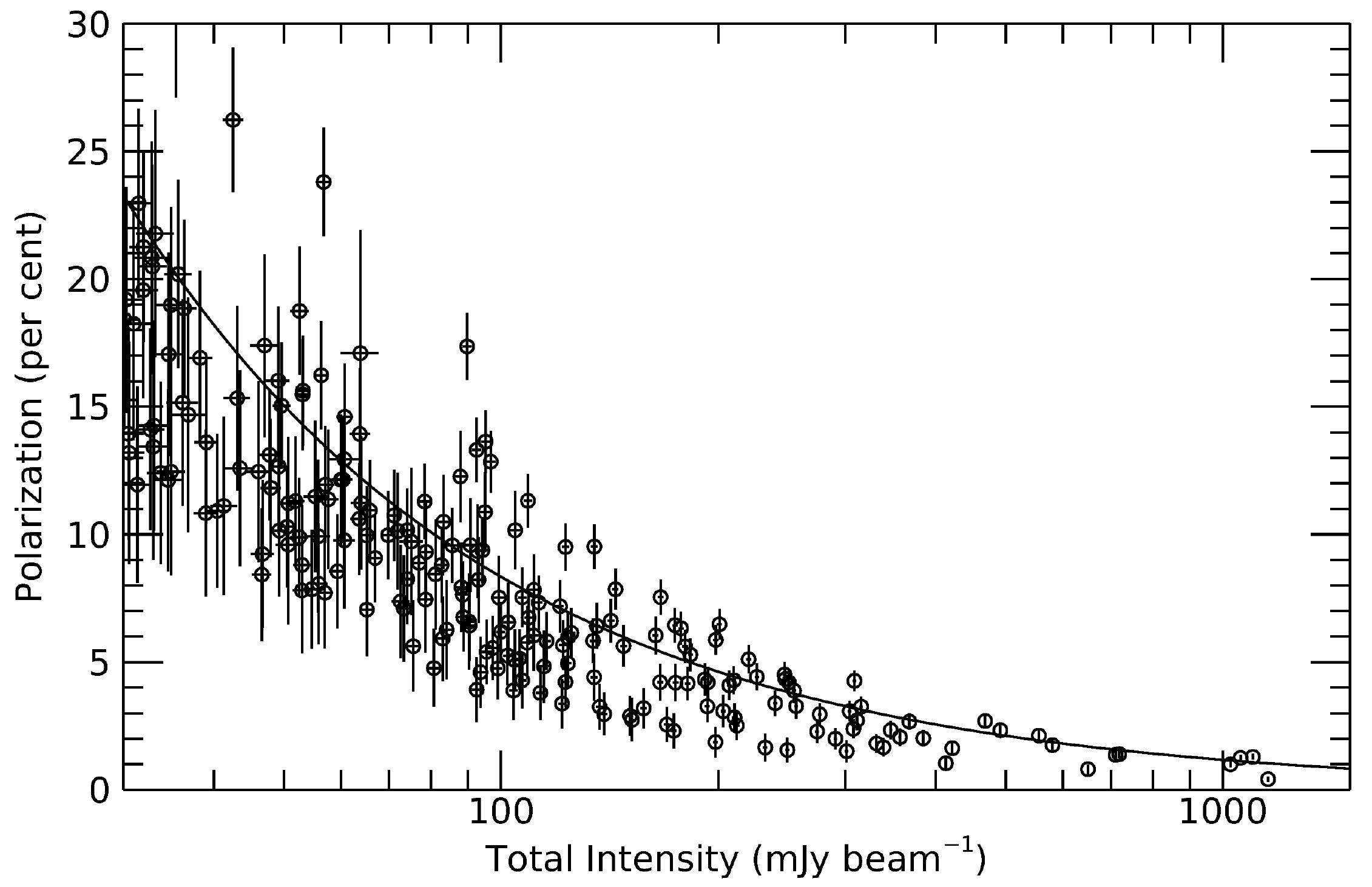}
\includegraphics[width=\columnwidth]{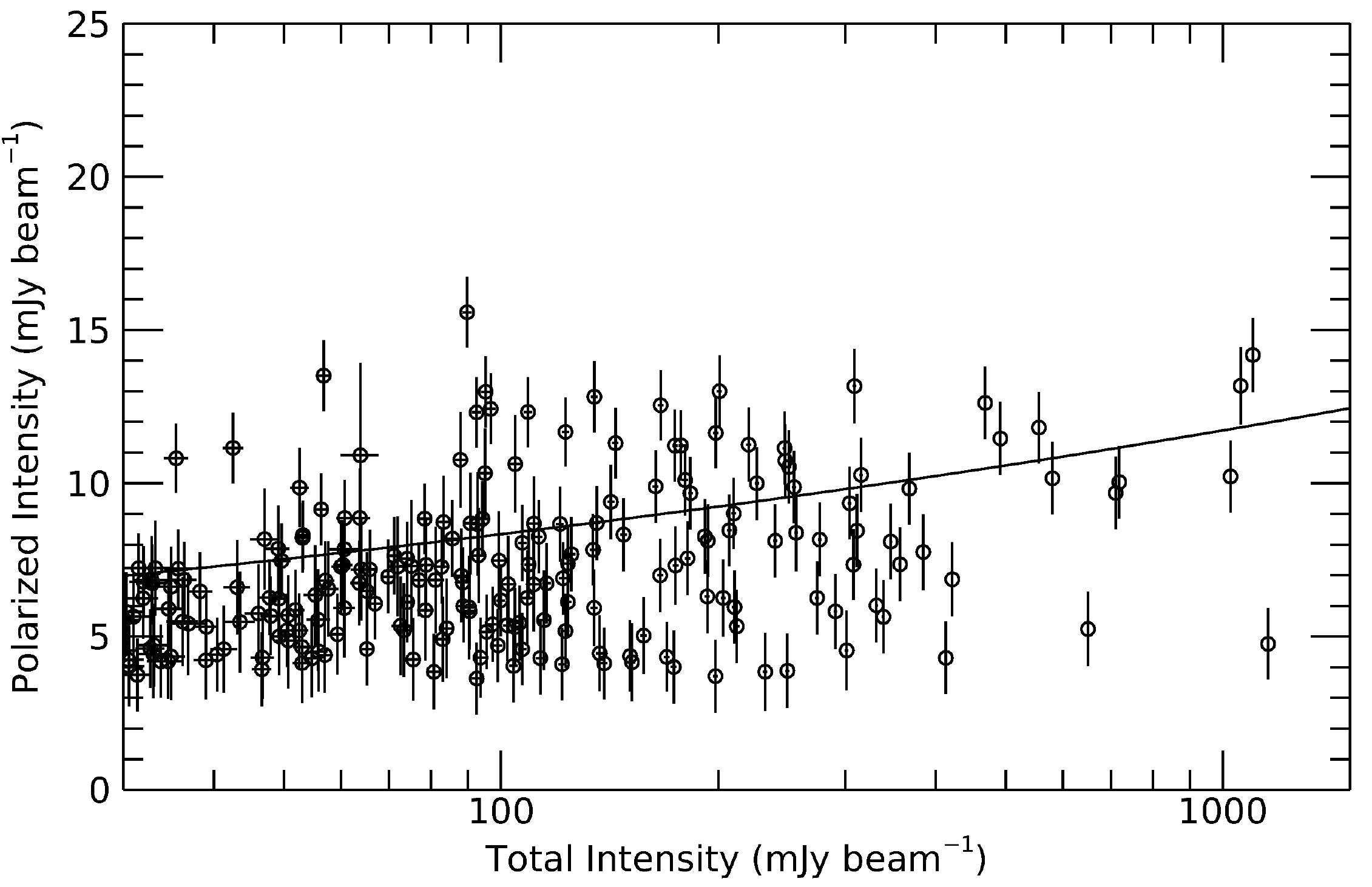}
\caption{Depolarization of POL-2 observations towards Perseus B1. Each point represents one of the polarization vectors shown in the left panel of Figure~\ref{fig:fig1_b1_polarization}. The vertical and horizontal lines show the uncertainties for the plotted parameters in each panel. \textit{Top}: De-biased polarization fraction $P$ as a function of the Stokes $I$ total intensity. \textit{Bottom}: De-biased polarized intensity $I_P$ as a function of the Stokes $I$ total intensity. The solid line in the top panel is the power-law fit (with index $\alpha \sim -0.85$) between the polarization fraction $P$ and the Stokes $I$ total intensity ($P \propto I^\alpha$, see Section~\ref{sub:pol2_perseusb1}). The solid line in the bottom panel is the same power-law fit as above, but multiplied by the Stokes~$I$ total intensity ($I_P \propto I^{\alpha+1}$).}
\label{fig:fig2_depolarization}
\end{figure}

The top panel of Figure~\ref{fig:fig2_depolarization} compares the fraction of polarization $P$ with the Stokes $I$ total intensity for each of the POL-2 vectors shown on the left panel of Figure~\ref{fig:fig1_b1_polarization}. There is a clear trend of decreasing fraction $P$ as a function of increasing Stokes $I$. If the total intensity is correlated with the column density \citep{Hildebrand1983}, this behavior can be understood as the result of a depolarization effect towards higher density regions of the cloud. The origin of this depolarization effect is discussed in Section~\ref{sec:discussion}. This trend does not mean, however, that the polarized intensity $I_P$ itself is decreasing. Indeed, the bottom panel of Figure~\ref{fig:fig2_depolarization} shows that $I_P$ may be in fact increasing slowly with Stokes $I$.

We fitted a power-law ($P \propto I^\alpha$) to the data in Figure~\ref{fig:fig2_depolarization} (top) using an error-weighted least-square minimization technique. We find a power index $\alpha = -0.85 \pm 0.01$, with a reduced chi-squared $\chi^2_r = 3.4$. This power-law is shown in both panels of Figure~\ref{fig:fig2_depolarization} as a solid line. The spread of data points relative to their uncertainties is responsible for the large $\chi^2_r$ value obtained, which indicates that fitting a single power-law may not be sufficient to account for the entire data set. The detailed effects of measurement uncertainties on the power-law fit between $P$ and $I$ are currently under investigation (K.~Pattle et al., in prep.).

The power index $\alpha \sim -0.85$ we find for B1 is nearly identical to the value measured in $\rho$~Ophiuchus~B by \citet{Soam2018} and relatively close to the index $\alpha \sim -0.8$ measured by \citet{Kwon2018} in $\rho$~Ophiuchus~A, both obtained from BISTRO data. Similarly, \citet{Matthews2002} previously found a power index $\alpha \sim -0.8$ in B1 using SCUPOL 850~$\mu$m measurements. The differences between POL-2 and SCUPOL polarization maps of B1 are quantified in Section~\ref{sub:scupol}.

\begin{figure}
	\includegraphics[width=\columnwidth]{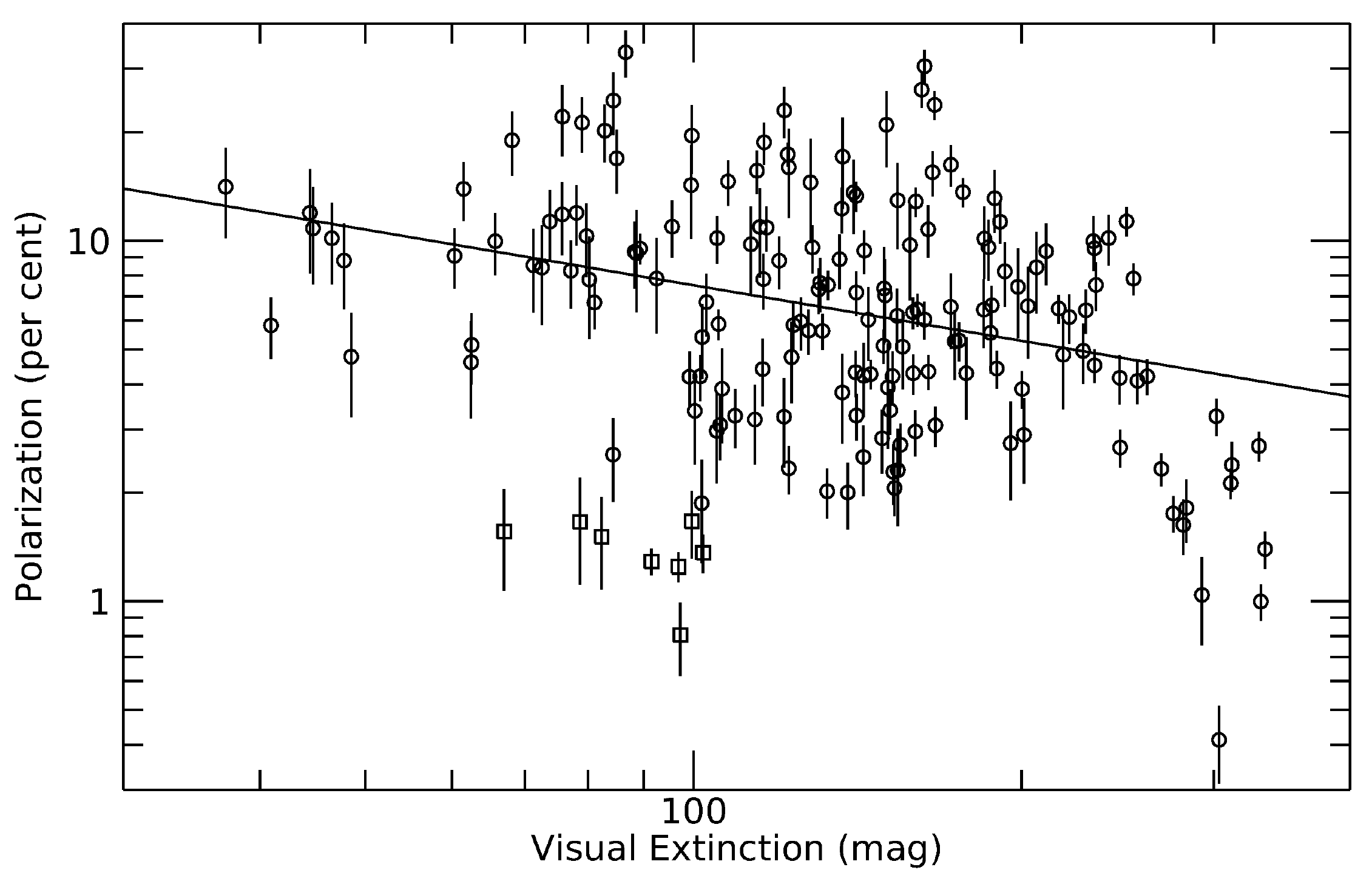}
	\caption{Relationship between the de-biased polarization fraction~$P$ and the visual extinction~$A_V$ in Perseus~B1. Each point represents one of the polarization vectors from the left panel of Figure~\ref{fig:fig1_b1_polarization} that also have \textit{Herschel}-derived opacity measurements. The visual extinction~$A_V$ is derived from the $300$~$\mu$m $\tau_{300}$ opacity map from \citet{Chen2016} assuming a reddening factor~$R_V=3.1$. The figure covers a range of extinction~$A_V$ from $30$~mag to $400$~mag. The vertical lines show the uncertainties for the polarization fraction~$P$. The $8$~polarization vectors found towards B1-c are identified with squares. The solid line is the power-law fit (with index $\beta \sim -0.5$) between the polarization fraction $P$ and the visual extinction~$A_V$ ($P \propto A_V^\beta$, see Section~\ref{sub:pol2_perseusb1}).}
	\label{fig:fig3_depolarization}
\end{figure}

However, in the context of grain alignment theory, it is more meaningful to take the optical depth into account when studying depolarization effects in molecular clouds. While an accurate modeling of the alignment efficiency of dust grains in Perseus~B1 will require a detailed analysis beyond the scope of this work, we can nonetheless begin to characterize the relationship between the polarized dust thermal emission and the visual extinction $A_V$ in the cloud by fitting a power-law of the form $P \propto A_V^\beta$ \citep[e.g.,][]{Alves2014}. Specifically, we know that the polarization fraction~$P$ of dust thermal emission obtained from submillimeter observations is proportional to the polarization efficiency $P_{ext}/A_V$ derived from measurements of the polarization fraction~$P_{ext}$ due to extinction at visible wavelengths \citep{Andersson2015}.

Figure~\ref{fig:fig3_depolarization} shows the relation between the polarization fraction~$P$ and the derived visual extinction~$A_V$ for the polarization vectors shown the left panel of Figure~\ref{fig:fig1_b1_polarization} that also have an associated opacity measurement in the $300$~$\mu$m $\tau_{300}$ opacity map from \citet{Chen2016}. We estimate the visual extinction~$A_V$ using Equation~A5 from \citet{Jones2015} and a version of the $\tau_{300}$ opacity map from \citet{Chen2016} that has been re-gridded from a pixel scale of $14$~arcsec to $12$~arcsec to match our observations. We also assume a reddening $R_V$ of $3.1$ which may be more representative of the diffuse interstellar medium \citep{Weingartner2001}, but should nonetheless serve as a reasonable lower limit for our estimation of the visual extinction $A_V$ across the cloud.

We fitted a power-law $P \propto A_V^\beta$ to the data shown in Figure~\ref{fig:fig3_depolarization} using an error-weighted least-square minimization technique. We find a power index $\beta = -0.51 \pm 0.03$, with a reduced chi-squared $\chi^2_r = 26.3$. This power-law is shown in Figure~\ref{fig:fig3_depolarization} as a solid line. The large reduced chi-squared $\chi^2_r$ value we find clearly indicates a poor fit to the data considering the spread of values and their uncertainties for the polarization fraction $P$ in Figure~\ref{fig:fig3_depolarization}. This could be explained in part by our use of a single reddening value to derive the visual extinction~$A_V$. Indeed, the reddening $R_V$ depends on the size distribution and composition of the dust grains, and so we do not expect this value to be constant across the cloud.

Nevertheless, the power index $\beta \sim -0.5$ we find in B1 is shallower than the power indices obtained from submillimeter observations in the Pipe-109 starless core ($\beta \sim -0.9$, \citealt{Alves2014, Alves2014corr}) and in the LDN~183 starless core ($\beta \sim -1.0$, \citealt{Andersson2015}). In fact, a power index $\beta \sim -0.5$ is closer to the power index $\beta \sim -0.6$ measured towards lower extinction regions ($A_V < 20$) of LDN~183 using visible and near-infrared observations \citep{Andersson2015}. Although Figure~\ref{fig:fig2_depolarization} clearly shows a depolarization effect with increasing total intensity~$I$, the power index $\beta \sim -0.5$ we find using the data in Figure~\ref{fig:fig3_depolarization} suggests that dust grains in Perseus~B1 are aligned more efficiently than in starless cores with comparable measures of visual extinction $A_V$. Since B1 is a site of on-going star formation, this may provide evidence that radiation from embedded young stellar objects can compensate for the expected loss of grain alignment with increasing visual extinction.

\subsection{Comparison with SCUPOL Legacy Data}
\label{sub:scupol}

\begin{figure}
\includegraphics[width=\columnwidth]{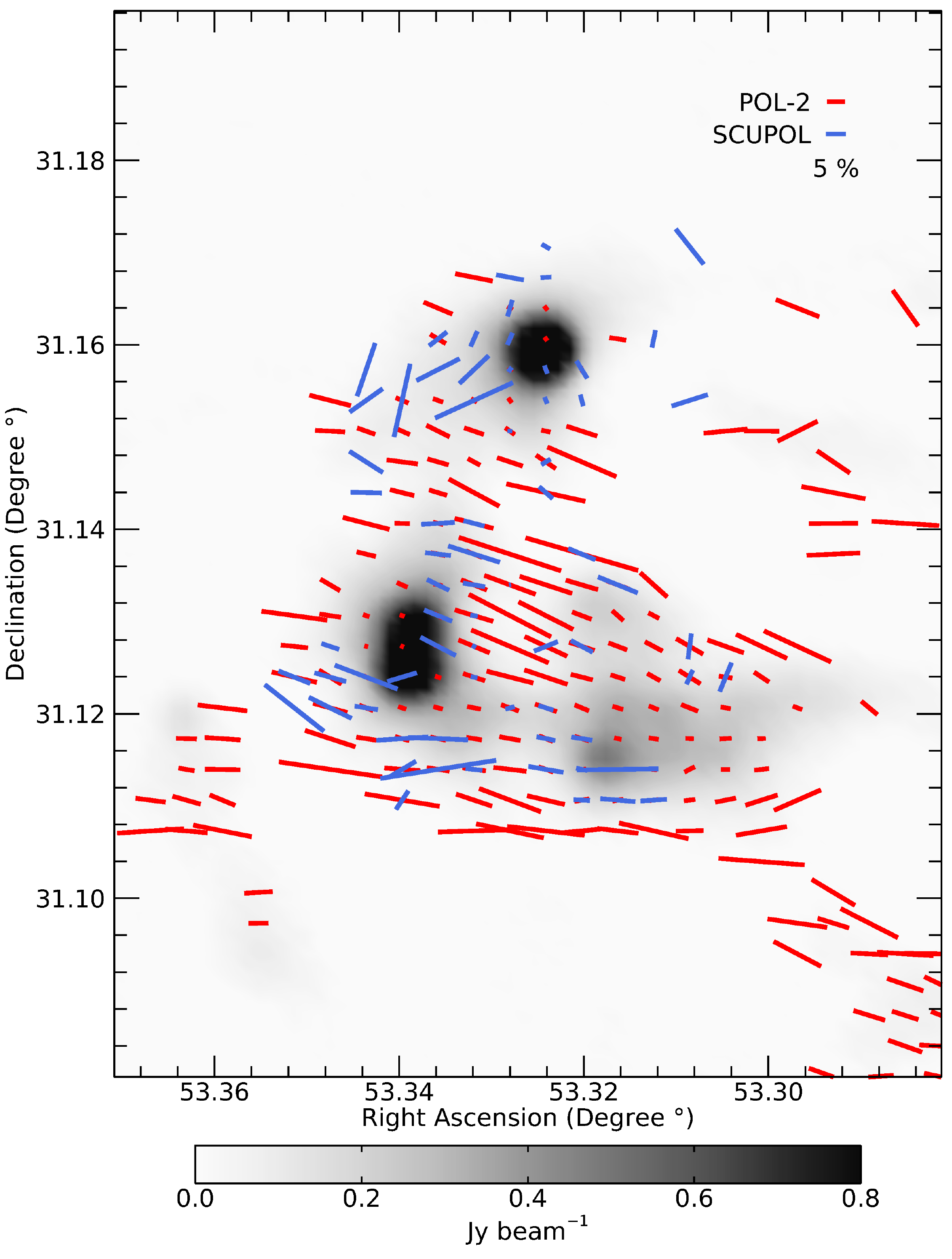}
\caption{Comparison of dust polarization at 850~$\mu$m between POL-2 (red) and SCUPOL (blue) towards Perseus B1. The gray scale indicates the Stokes $I$ total intensity measured with POL-2. The length of each vector is determined by its associated polarization fraction $P$ (per cent). The SCUPOL polarization vectors from \citet{Matthews2009} have been re-binned to match the exact position and pixel scale (from 10~arcsec to 12~arcsec) of the POL-2 observations.}
\label{fig:fig4_scupol}
\end{figure}

As mentioned in Section~\ref{sub:scuba2}, Perseus~B1 was previously observed at 850~$\mu$m with the SCUPOL polarimeter \citep{Matthews2002}. Here we specifically compare the BISTRO results presented in Section~\ref{sub:pol2_perseusb1} to the polarization data of B1 found in the SCUPOL Legacy Catalog \citep{Matthews2009}.

Figure~\ref{fig:fig4_scupol} compares the BISTRO observations to their equivalent data set in the SCUPOL Legacy Catalog, with the POL-2 polarization vectors (same as Figure~\ref{fig:fig1_b1_polarization}) in red and the SCUPOL vectors in blue. To have a significant number of SCUPOL vectors for this comparison, we relaxed their selection criteria compared to POL-2. For the SCUPOL data, we use $I/\sigma_I\!>\!2$, $P/\sigma_P\!>\!2$, and $\sigma_P\!<\!10$~per cent. These relaxed criteria provide a total catalog of 69~vectors, compared to only 17 when applying the same selection criteria as for the POL-2 data.

At best, the relaxed catalog of SCUPOL vectors achieves a sensitivity of $0.5$ per cent in polarization fraction and an uncertainty of $5.5^\circ$ in polarization angle, with mean values for $\sigma_P$ of $2.7$ per cent and for $\sigma_\Phi$ of $10.3^\circ$.

\begin{figure}
\includegraphics[width=\columnwidth]{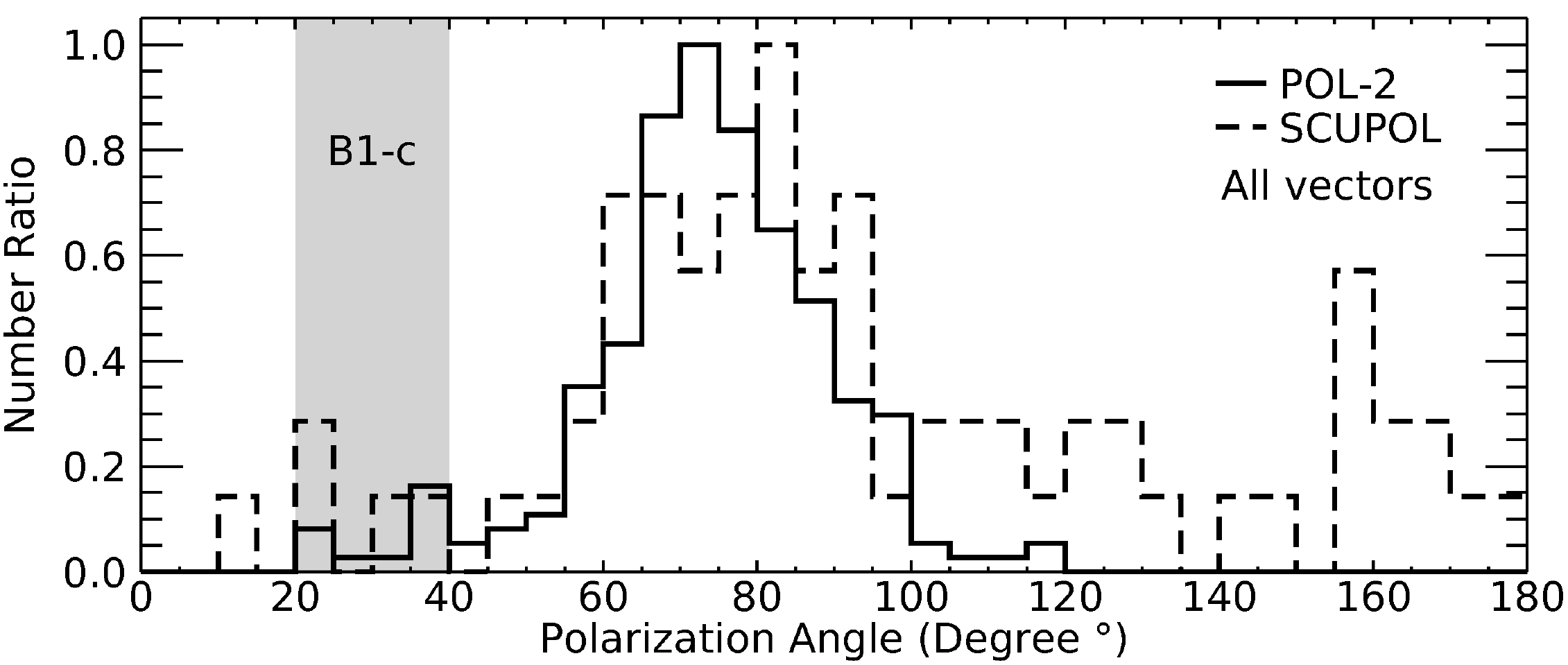}
\includegraphics[width=\columnwidth]{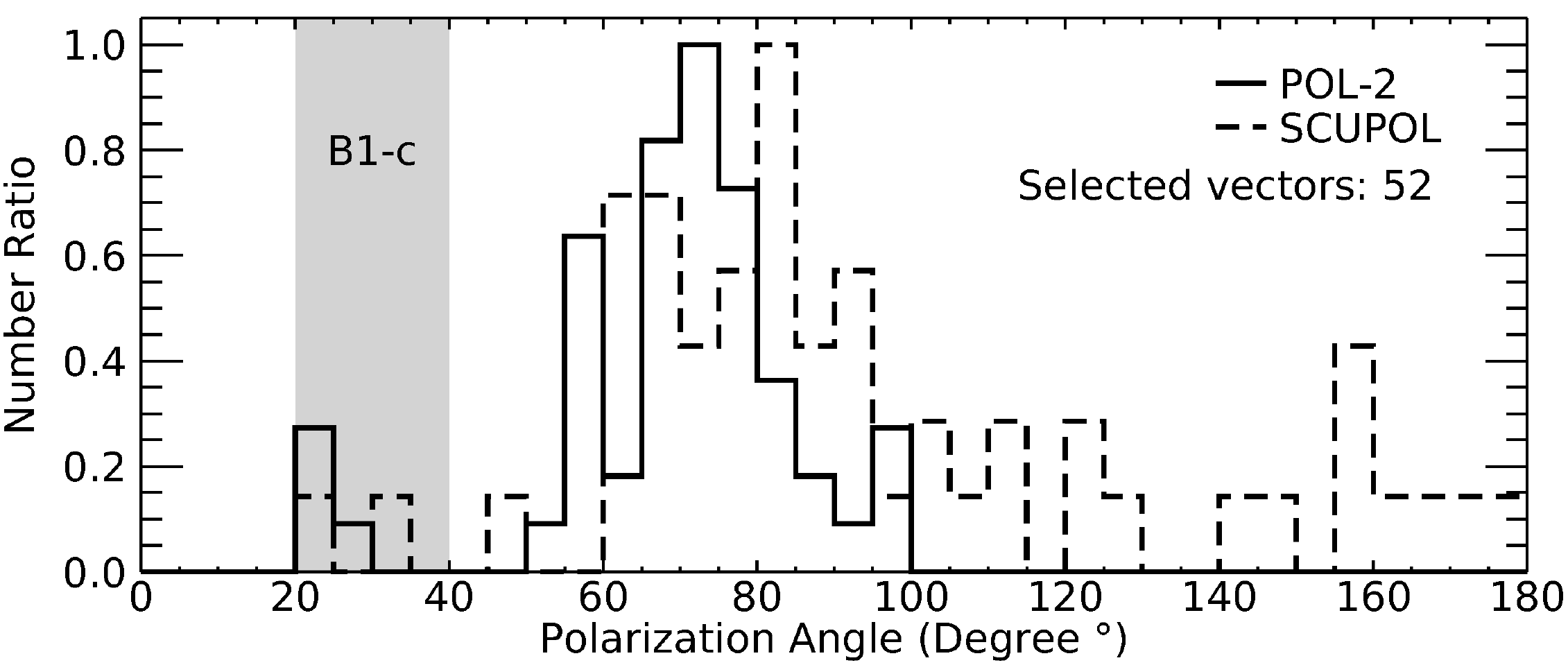}
\caption{Histograms of polarization angles for Perseus B1 from POL-2 and SCUPOL. The number of vectors in each bin is normalized by the maximum value of the histogram ($N_{\text{bin}} / N_{\text{max}}$) for a given sample of polarization angles. \textit{Top}: Histogram including all the POL-2 (224) and SCUPOL (69) polarization vectors shown in Figure~\ref{fig:fig1_b1_polarization} and Figure~\ref{fig:fig4_scupol} respectively. \textit{Bottom}: Histogram including only the 52 positions for which there exists both a POL-2 and a SCUPOL polarization vector in Figure~\ref{fig:fig4_scupol}. In both panels, the range of polarization angles associated with the protostellar source B1-c is shown in gray.}
\label{fig:fig5_scupol_histo}
\end{figure}

Figure~\ref{fig:fig5_scupol_histo} shows the distribution of angles for both the POL-2 and SCUPOL polarization maps. The top panel shows the histogram including all the POL-2 and SCUPOL polarization vectors shown in Figure~\ref{fig:fig4_scupol}, normalized by the maximum value in each distribution. Both distributions peak between 65$^\circ$ and 85$^\circ$. The bottom panel shows the normalized distributions only for those vector positions that are common (i.e., spatially overlapping within the same pixel) to both SCUPOL and POL-2. There are 52~such positions in the maps. 

We used a Kolmogorov-Smirnov test to compare the distributions shown at the bottom of Figure~\ref{fig:fig5_scupol_histo}. Specifically, a two-sample Kolmogorov-Smirnov test provides the probability that two independent data samples are drawn from the same intrinsic distribution by measuring the maximum distance between the cumulative probability distribution of each sample. For example, if both the SCUPOL and POL-2 values for the selected co-spatial vectors were exact measurements of the $850$~$\mu$m polarization towards Perseus~B1, then we would expect the two catalogs of polarization angles, and therefore their respective cumulative probability distributions, to be identical and the Kolmogorov-Smirnov test to return a $100$~per cent probability that they are drawn from the same intrinsic distribution of polarization angles. In reality, the POL-2 and SCUPOL distributions shown in the bottom panel of Figure~\ref{fig:fig5_scupol_histo} are not identical even though they probe the same positions in B1, and so the Kolmogorov-Smirnov test becomes a way of quantifying the difference between them since it makes no assumption about the nature of the aforementioned intrinsic distribution.

In this case, we find a low likelihood ($0.6$~per cent) that both POL-2 and SCUPOL distributions in the bottom panel of Figure~\ref{fig:fig5_scupol_histo} are drawn from the same intrinsic distribution of polarization angles (with a maximum deviation $D = 0.39$ between the cumulative probability distributions). In other words, based only on the $52$~available co-spatial vectors in each sample, a two-sample Kolmogorov-Smirnov test shows that the distributions of POL-2 and SCUPOL polarization angles are significantly different from each other. If we set the selection criteria for POL-2 vectors to be identical to those applied for SCUPOL vectors, we find instead 64~positions with vectors common to both catalogs. This relaxed data set does not, however, improve the results of the Kolmogorov-Smirnov test. 

\begin{figure}
\includegraphics[width=\columnwidth]{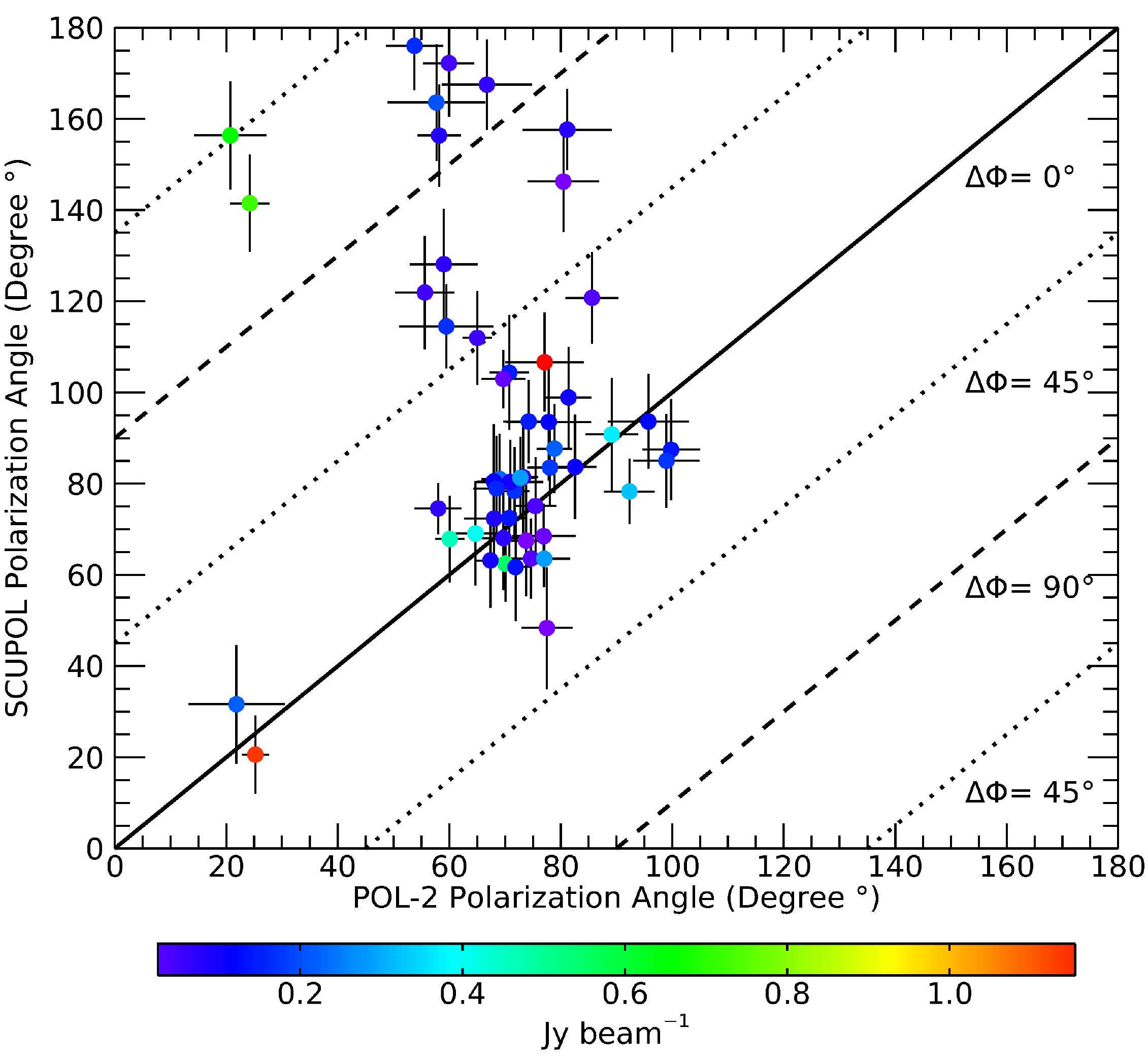}
\includegraphics[width=\columnwidth]{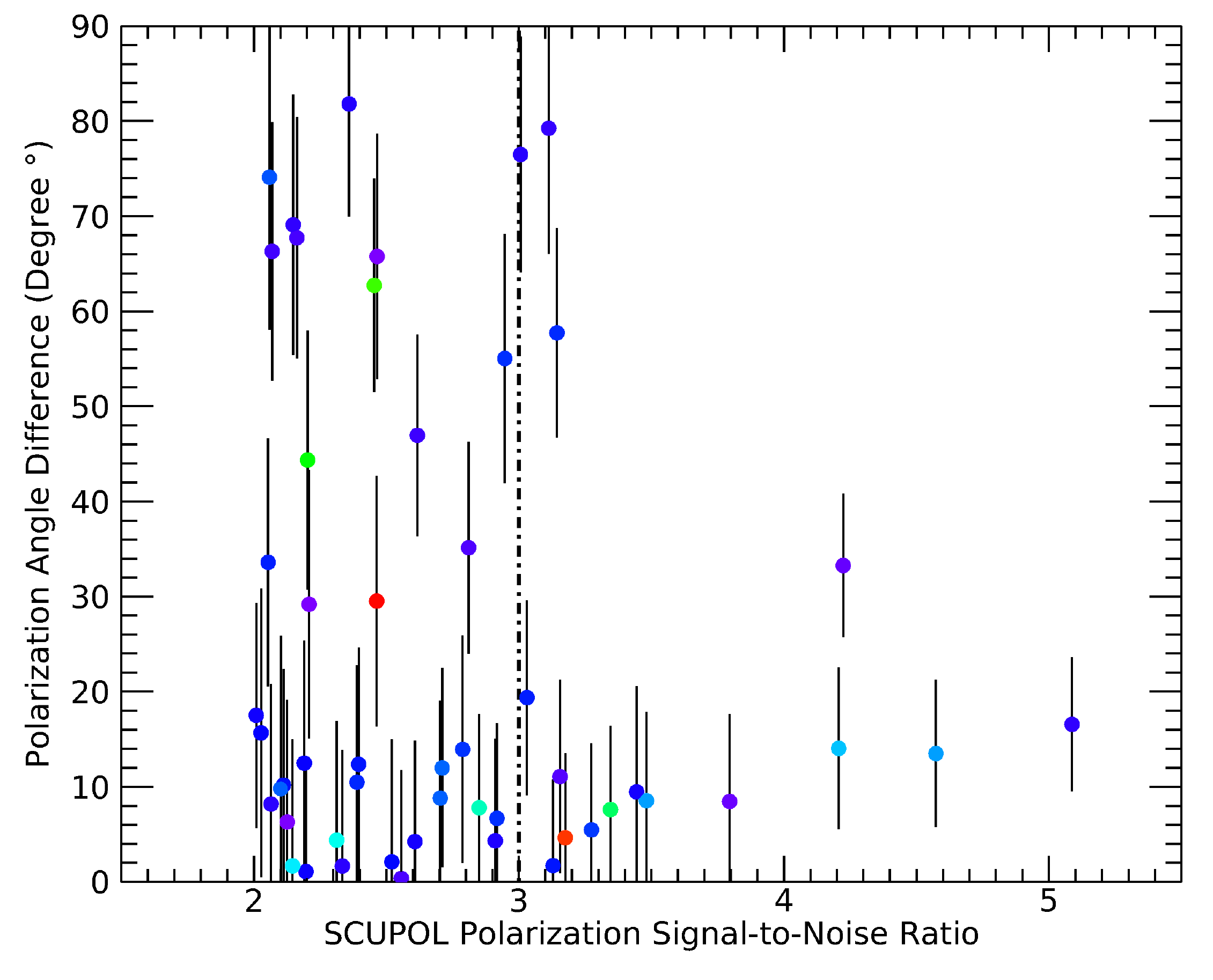}
\caption{\textit{Top}: Comparison of polarization angles for the 52 pairs of spatially overlapping POL-2 and SCUPOL vectors plotted in Figure~\ref{fig:fig4_scupol}. The plain line follows the 1:1 correspondence, and the dotted and dashed lines respectively trace differences of a 45~degrees and 90~degrees in polarization angle. \textit{Bottom}: Difference of polarization angle between each pair of POL-2 and SCUPOL vector ($\Delta\Phi = \left| \Phi_{\text{SCUPOL}} - \Phi_{\text{POL-2}} \right|$) as a function of the signal-to-noise ratio (SNR) of the polarization fraction measured with SCUPOL ($P_{\text{SCUPOL}}/\sigma_{P_{\text{SCUPOL}}}$). The vertical dashed line indicates a SNR of~3. In both panels, the color scale indicates the Stokes~$I$ intensity of the POL-2 vector associated with each point.}
\label{fig:fig6_scupol_comparison}
\end{figure}

Figure~\ref{fig:fig6_scupol_comparison} expands the comparison shown in Figure~\ref{fig:fig5_scupol_histo} (bottom) between the POL-2 and SCUPOL polarization angles for pairs of spatially overlapping vectors. The top panel of Figure~\ref{fig:fig6_scupol_comparison} shows that most outliers from the 1:1 correspondence line are found towards lower intensity regions ($I < 200$~mJy~beam$^{-1}$), as measured from POL-2 Stokes~$I$. Furthermore, in Figure~\ref{fig:fig6_scupol_comparison} (bottom), the vector pairs displaying the largest angular difference ($\left| \Phi_{\text{SCUPOL}} - \Phi_{\text{POL-2}} \right|$) are found near or below a SNR of~3 for the polarization fraction ($P_{\text{SCUPOL}}/\sigma_{P_{\text{SCUPOL}}} \lesssim 3$) measured with SCUPOL. Although the pairs of vectors at high SNR ($P_{\text{SCUPOL}}/\sigma_{P_{\text{SCUPOL}}} > 4$) also exhibit a non-negligible angular difference, this effect is not nearly as pronounced as for the low SNR vectors ($P_{\text{SCUPOL}}/\sigma_{P_{\text{SCUPOL}}} \lesssim 3$). This disparity between POL-2 and SCUPOL could therefore be explained by the relatively high noise levels in the SCUPOL Legacy data. 

\section{Analysis}
\label{sec:analysis}

\subsection{Angular Dispersion Analysis and Davis-Chandrasekhar-Fermi Method}
\label{sub:dcf}

The magnetic field strength in molecular clouds can be estimated through the Davis-Chandrasekhar-Fermi (DCF) method \citep{Davis1951, CF1953}. This technique relies on the assumption that turbulent motions in the gas will locally inject randomness in the observed morphology of a large-scale magnetic field. Since polarization vectors are expected to trace the plane-of-sky component of the magnetic field, we can infer the strength of this component by measuring the dispersion of polarization angles relative to the large-scale field orientation. This technique, however, also requires the velocity dispersion and the density of the gas in the cloud to be known beforehand.

According to \citet{Crutcher2004}, the DCF equation for the plane-of-sky magnetic field strength $B_{\text{pos}}$ can be written as: 
\begin{equation}
B_{\text{pos}} = A \; \sqrt[]{4 \pi \rho} \; \frac{\delta V}{\delta \Phi} \, ,
\label{eq:dcf}
\end{equation}
where $\rho$ is the density, $\delta V$ is the velocity dispersion of the gas in the cloud, $\delta \Phi$ is the dispersion of polarization angles (in radians), and $A$ is a correction factor usually assumed to be $\sim 0.5$. The correction factor $A$ is included to account for the three-dimensional nature of the interplay between turbulence and magnetism \citep[e.g.,][]{Ostriker2001}. There is, however, a caveat to Equation \ref{eq:dcf}, namely that it cannot intrinsically account for changes in the large-scale field morphology. As a consequence, the technique from \citet{Crutcher2004} was modified by \citet{Pattle2017} to take large-scale variations in field morphology into account when calculating the magnetic field strength in Orion A.

Specifically, \citet{Pattle2017} calculate the dispersion $\delta \Phi$ of polarization angles in Equation~\ref{eq:dcf} with an unsharp-masking technique. First, the large-scale component of the field is found by smoothing the map of polarization angles using $3 \times 3$-pixels boxes. This smoothed map is then subtracted from the original to obtain a map of the residual polarization angles. Finally, the dispersion $\delta \Phi$ is obtained from the mean value of the residual angles fitting a specific set of conditions. This approach therefore cancels the contribution of a changing field morphology to the dispersion of polarization angles at scales larger than the smoothed mean-field map.

In our work, we instead apply the improved DCF method developed by \citet{Hildebrand2009} and \citet{Houde2009}, which was also adapted for polarimetric data obtained by interferometers such as the SMA and CARMA \citep{Houde2011, Houde2016}. This technique avoids the problem of spatial changes in field morphology by using an angular dispersion function (sometimes called structure function) rather than the dispersion of polarization angles around a mean value. Furthermore, the angular dispersion technique from \citet{Houde2009} was independently tested using both R-band \cite[e.g.,][]{Franco2010} and submillimeter \cite[e.g.,][]{Ching2017} polarimetric observations to characterize the magnetic and turbulent properties of star-forming regions.

This angular dispersion function is calculated by taking the angular difference between every pair of polarization vectors in a given map as a function of the distance between them. This technique effectively traces the ratio between turbulent and magnetic energies, which can then be fitted without any prior assumptions on the turbulence in the cloud or the morphology of the large-scale field \citep{Hildebrand2009}. As before, this analysis can be used to estimate the strength of the plane-of-sky magnetic field component if the density and velocity dispersion of the cloud are known. Additionally, it can be used to measure the effect of integrating turbulent cells along the line-of-sight within a telescope beam, effectively constraining the theoretical factor $A$ included in Equation~\ref{eq:dcf} \citep{Houde2009}.

We first need to define the relevant quantities for the dispersion analysis presented in this paper. The difference in polarization angle between two vectors as a function of distance $\ell$ is defined as: $\Delta\Phi(\ell) \equiv \Phi(\textbf{x}) - \Phi(\textbf{x} + \bm{\ell})$, where $\Phi(\textbf{x})$ is the angle $\Phi$ of the polarization vector found at a position $\textbf{x}$ in the map and $\bm{\ell}$ is the angular displacement between two vectors. With this quantity, we can define the angular dispersion function as formulated by \citet{Houde2009}:
\begin{equation}
1- \left\langle \cos[\Delta\Phi(\ell)] \right\rangle \, ,
\label{eq:dispersion}
\end{equation}
where $\left\langle ... \right\rangle$ is the average over every pair of vectors separated by a distance $\ell$. Since Equation~\ref{eq:dispersion} is essentially a measure of the mean difference in polarization angles as a function of distance, it is accurate to describe it as an angular dispersion function.

The magnetic field $\textbf{B}(\textbf{x})$ in the cloud at a position $\textbf{x}$ can be written as a combination of a large-scale (or ordered) component $\textbf{B}_{o}(\textbf{x})$ and a turbulent component $\textbf{B}_{t}(\textbf{x})$, i.e., $\textbf{B}(\textbf{x}) = \textbf{B}_{o}(\textbf{x})+\textbf{B}_{t}(\textbf{x})$. Furthermore, we define the ratio between the average energy of the turbulent component to that of the large-scale component as $\left\langle B_t^2 \right\rangle / \left\langle B_o^2 \right\rangle$ and the ratio between the average energy of the turbulent component to that of the total magnetic field as $\left\langle B_t^2 \right\rangle / \left\langle B^2 \right\rangle$. Both quantities can be obtained from fitting the angular dispersion function.

To relate the magnetic fields and turbulence, we also need to define the turbulent properties of the cloud. Specifically, we require the number $N$ of independent magnetic turbulent cells observed for a column of dust along the line-of-sight and within a telescope beam from:
\begin{equation}
N = \Delta' \, \frac{\left( \delta^2 + 2 W^2 \right)}{\sqrt[]{2 \pi}\, \delta^3} \, ,
\label{eq:turbulence}
\end{equation}
where $\delta$ is the turbulent correlation length scale of the magnetic field, $W$ is the radius of the circular telescope beam (specifically, $\text{FWHM} = 2 \; \sqrt[]{2 \, \text{ln}2} \, W$), and $\Delta'$ is the effective thickness of the cloud \citep[see Equation~52 in][]{Houde2009}. The turbulent correlation length scale $\delta$ can be understood as the typical size of a magnetized turbulent cell in the cloud. In this specific case, the turbulence is supposedly isotropic and the turbulent correlation length scale $\delta$ is assumed to be smaller than the thickness $\Delta'$ of the cloud.

If the physical depth of the cloud is not known beforehand, the effective thickness $\Delta'$ can be estimated from the autocorrelation function of the integrated polarized intensity across the cloud \citep[see Equation~51 in][]{Houde2009}. This autocorrelation function is defined as:
\begin{equation}
\left\langle I_P^2(\ell) \right\rangle \equiv \left\langle I_P(\textbf{x})  \, I_P(\textbf{x} + \bm{\ell}) \right\rangle\, ,
\label{eq:pi_autocorrelation}
\end{equation}
from which we use the width at half-maximum to evaluate $\Delta'$. This approach, however, assumes that the spatial distribution of polarized dust emission on the plane-of-sky is an adequate probe of the cloud's properties along the line-of-sight, which we believe to be reasonable in the case of dense molecular clouds.

The detailed derivations given by \citet{Hildebrand2009} and \citet{Houde2009} show that the relationship between the angular dispersion function and the magnetic and turbulent properties of a molecular cloud can be expressed by the following equation:
\begin{equation}
1- \left\langle \cos[\Delta\Phi(\ell)] \right\rangle \simeq \frac{1}{N} \, \frac{\left\langle B_t^2 \right\rangle}{\left\langle B_o^2 \right\rangle} - b^2(\ell) + a \, \ell^2 \, ,
\label{eq:adf}
\end{equation}
where $a$ is the first Taylor coefficient of the ordered autocorrelation function, and $b^2(\ell)$ is the autocorrelated turbulent component of the dispersion function \cite[see Equations~53 and 55 in][]{Houde2009}. Specifically, the Taylor coefficient $a$ is related to the large-scale structure of the magnetic field. Additionally, we can write this autocorrelated turbulent component as: 
\begin{equation}
b^2(\ell) = \frac{1}{N} \, \frac{\left\langle B_t^2 \right\rangle}{\left\langle B_o^2 \right\rangle} \, e^{-\ell^2 / 2(\delta^2+2W^2)} \, .
\label{eq:autocorrelation}
\end{equation}
Since the beam radius $W$ and the effective cloud thickness $\Delta'$ can be considered as known quantities, we only need to fit three parameters to the angular dispersion function: the ratio of turbulent energy to large-scale magnetic energy $\left\langle B_t^2 \right\rangle / \left\langle B_o^2 \right\rangle$, the turbulent correlation length scale $\delta$ of the magnetic field, and the first Taylor coefficient $a$ of the ordered autocorrelation function.

Finally, \citet{Houde2009} rewrote the DCF equation (see Equation \ref{eq:dcf}) for the plane-of-sky strength of the magnetic field to calculate it directly from the ratio of turbulent energy to total magnetic energy $\left\langle B_t^2 \right\rangle / \left\langle B^2 \right\rangle$ in the cloud. This new formulation of the DCF equation can be written as: 
\begin{equation}
B_{\text{pos}} \simeq \sqrt[]{4 \pi \rho} \: \delta V \, \left[  \frac{\left\langle B_t^2 \right\rangle}{\left\langle B^2 \right\rangle} \right]^{-1/2} \, ,
\label{eq:dcf_houde}
\end{equation}
where as previously $\rho$ is the density and $\delta V$ is the one-dimensional velocity dispersion for the gas (see Equation~57 in \citealt{Houde2009} and Equation~26 in \citealt{Houde2016}). The gas density $\rho$ takes the form $\rho~=~\mu \, m_{H} \, n(\text{H}_2)$, where $\mu = 2.8$ is the mean molecular weight of the gas \citep{Kauffmann2008}, $m_{H}$ is the mass of an hydrogen atom, and $n(\text{H}_2)$ is the number density of hydrogen molecules in the cloud.

Once the strength of the plane-of-sky component of the magnetic field has been calculated with Equation~\ref{eq:dcf_houde}, it becomes possible to evaluate the magnetic critical ratio $\lambda_c$ of the studied molecular cloud \citep{Crutcher2004}. The critical ratio $\lambda_c$ can be estimated from the plane-of-sky amplitude of the magnetic field with the following equation:
\begin{equation}
\lambda_c \simeq 7.6 \times 10^{-21} \, \frac{N(\text{H}_2)}{B_{\text{pos}}} \, ,
\label{eq:crit_ratio}
\end{equation}
where $N(\text{H}_2)$ is the typical column density of molecular hydrogen in the cloud. If $\lambda_c < 1$, then the molecular cloud is magnetically subcritical and the magnetic field is sufficiently strong to stop its gravitational collapse. If $\lambda_c > 1$, the cloud is instead magnetically supercritical and the magnetic field alone cannot support the cloud against its self-gravity.

\subsection{Cloud Characteristics and Magnetic Field Strength in Perseus~B1}
\label{sub:dispersion_results}

\begin{figure}
\includegraphics[width=\columnwidth]{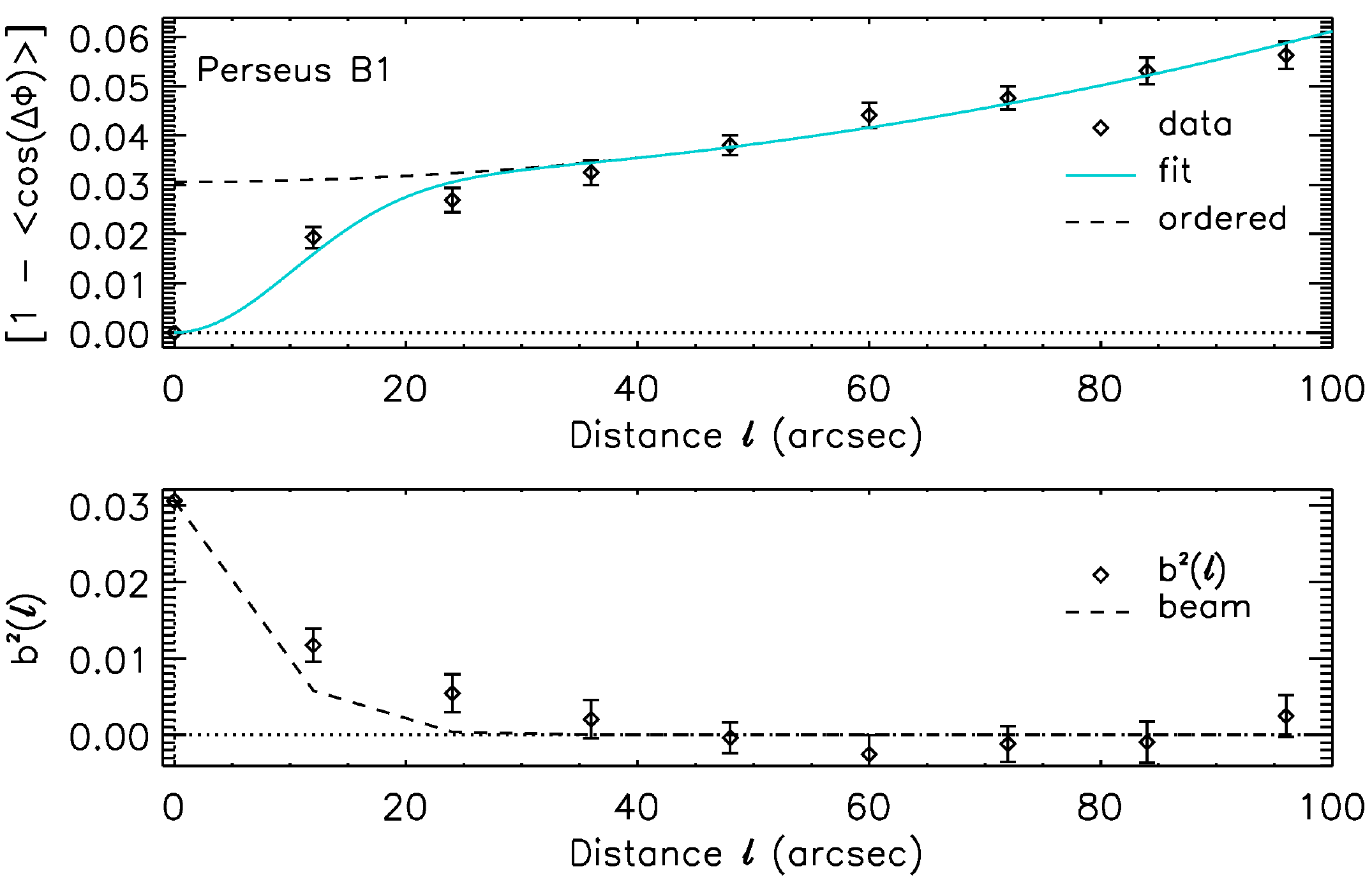}
\caption{Dispersion of polarization angles for POL-2 observations of Perseus B1. \textit{Top}: The angular dispersion function $[1-\text{cos}(\Delta \Phi)]$ as a function of the distance $\ell$. The fit of Equation \ref{eq:adf} to the data is shown with (blue solid line) and without (black dashed line) including the autocorrelation function $b^2(\ell)$ defined in Equation \ref{eq:autocorrelation}. \textit{Bottom}: Signal-integrated turbulence autocorrelation function $b^2(\ell)$ as a function of distance $\ell$. The black dashed line shows the contribution of the telescope beam alone.}
\label{fig:fig7_dcf}
\end{figure}

Following Section~\ref{sub:dcf}, we determine the angular dispersion function from the POL-2 data of Perseus B1. We include in this analysis all the POL-2 polarization vectors found in a 240~arcsec-wide square centered on the position (03$^{\text{h}}$ 33$^{\text{m}}$ 20$^{\text{s}}$.45, $+$31$^{\circ}$ 07$'$ 50$''$.16), as illustrated in the right panel of Figure~\ref{fig:fig1_b1_polarization}. This region covers most of the embedded young stellar objects in the densest parts of Perseus B1. The resulting angular dispersion function is shown in the top panel of Figure~\ref{fig:fig7_dcf} as a function the distance $\ell$ in bins of 12~arcsec. The observed steady increase of this function with $\ell$ at small spatial scales (0.01 to 1.0~pc) is also a behavior seen in other studies using this technique \citep[e.g.,][]{Houde2009,Houde2016,Franco2010,Ching2017,Chuss2019}.

The angular dispersion function was fitted with Equation~\ref{eq:adf} to obtain $\delta$ and $\left\langle B_t^2 \right\rangle / \left\langle B_o^2 \right\rangle$ using an effective cloud depth $\Delta'$ of $84$~arcsec, and a beam radius $W$ of $6.2$~arcsec (or a FWHM of $14.6$~arcsec) at 850~$\mu$m. The reduced chi-squared value for this fit is $\chi_r^2=1.5$. The results of the fit to the angular dispersion, including $\left\langle B_t^2 \right\rangle / \left\langle B^2 \right\rangle$, are given in Table~\ref{tab:houde_results}. Additionally, the resulting turbulent autocorrelation function $b^2(\ell)$ is shown on the bottom panel of Figure~\ref{fig:fig7_dcf}. 

At a distance of 295~pc \citep{Ortiz2018}, the effective cloud depth $\Delta'$ of $84$~arcsec in B1 represents a physical depth of $\sim~0.1$~pc. While this effective cloud depth $\Delta' \sim 0.1$~pc was derived independently from the autocorrelation function of the polarized intensity $I_P$ (see Section~\ref{sub:dcf}), it is nonetheless comparable to the typical width of dense filaments in star-forming regions \citep[e.g.,][]{Arzoumanian2011, Andre2014, Koch2015, Andre2016}. For reference, the square region shown in the right panel of Figure~\ref{fig:fig1_b1_polarization} has a width $\sim 0.4$~pc ($\sim 270$~arcsec).

The exact distance to the Perseus molecular cloud, and to B1 in particular, is still subject to some ambiguity. Indeed, different methods  provide a wide range of values from 235~pc \citep[22~GHz water maser parallaxes;][]{Hirota2008, Hirota2011} to 315~pc \citep[photometric reddening;][]{Schlafly2014}. Furthermore, \citet{Schlafly2014} found a gradient of distances from the western (260~pc) to the eastern (315~pc) parts of the Perseus molecular cloud complex. However, recent parallaxes measurements with the \textit{Gaia} space telescope instead suggest a smaller range of distances between NGC~1333 (295~pc) and IC~348 (320~pc) \citep{Ortiz2018}. According to these \textit{Gaia} results, the distance to B1 is similar to that of NGC~1333 at 295~pc. This distance to B1 assumes that the young stellar objects used for these parallaxes measurements provide a good estimate of the clump's true position along the line-of-sight.

Perseus~B1 was mapped in emission from several NH$_3$ inversion transitions at $\sim$24~GHz by GAS (the first data release of the survey was presented by \citealt{Friesen2017}). NH$_3$ is a commonly-used selective tracer of moderately dense gas ($n \gtrsim$ a few $10^3$~cm$^{-3}$; \citealt{Shirley2015}). The NH$_3$ (1,1) emission closely follows the intensity detected with POL-2 across the cloud (GAS Consortium, in prep.). The velocity dispersion of the gas along each line-of-sight was obtained through simultaneous modeling of hyperfine structure of the detected NH$_3$ (1,1) and (2,2) inversion line emission. Assuming that the (1,1) and (2,2) lines share the same line-of-sight velocity, velocity dispersion, and excitation temperature, the analysis produces maps of the aforementioned parameters along with the gas kinetic temperature, and the total column density of NH$_3$. Further details of the modeling are given in \citet{Friesen2017}.

For the region delimited by the square in the right panel of Figure~\ref{fig:fig1_b1_polarization}, we find an average velocity dispersion $\delta V = 0.29$~km~s$^{-1}$, with a standard deviation $\sigma_{\delta V} = 0.11$~km~s$^{-1}$. The uncertainties for individual line width measurements are typically $< 0.05$~km~s$^{-1}$. We therefore use the velocity dispersion $\delta V = (2.9 \pm 1.1) \times 10^4$~cm~s$^{-1}$ to calculate the plane-of-sky amplitude of the magnetic field with Equation~\ref{eq:dcf_houde}.

The number density $n(\text{H}_2)$ of the gas in Perseus~B1 is also calculated from the same GAS NH$_3$ data (\citealt{Friesen2017}; GAS Consortium, in prep.). Specifically, we follow the relation described by \citet{Ho1983} between density, excitation temperature, and gas kinetic temperature to estimate the number density $n(\text{H}_2)$ in B1, assuming the NH$_3$ emission in B1 can be approximated by a two-level system. First, for the denser regions associated with polarized emission, we find a mean gas temperature of $11.6$~K with a standard deviation of $1.2$~K, and a mean excitation temperature of $6.5$~K with a standard deviation of $0.4$~K. Using these temperatures, we calculate a mean density $n(\text{H}_2) = (1.5 \pm 0.3) \times 10^5$~cm$^{-3}$. If the typical depth of the dense material in B1 is indeed $\sim 0.1$~pc, we then find a column density $N(\text{H}_2) = (4.7 \pm 0.9) \times 10^{22}$~cm$^{-2}$ in agreement with the values obtained from fitting far-infrared and submillimeter measurements of dust thermal emission \citep{Sadavoy2013,Chen2016}. Finally, assuming a molecular weight $\mu = 2.8$ \citep{Kauffmann2008}, we derive an average gas density $\rho = (7.0 \pm 1.4) \times 10^{-19}$~g~cm$^{-3}$. 

The ratio $\left\langle B_t^2 \right\rangle / \left\langle B^2 \right\rangle$ of turbulent-to-total magnetic energy given in Table~\ref{tab:houde_results} can be used to calculate the plane-of-sky strength of the magnetic field in Perseus B1 using Equation~\ref{eq:dcf_houde}. Combined with the values given previously for the density $\rho$ and velocity dispersion $\delta V$, we calculate the plane-of-sky strength of the magnetic field in Perseus B1 to be $120 \pm 60$~$\mu$G.

We compare the plane-of-sky strength of the magnetic field derived from the angular dispersion analysis \citep{Houde2009} with the one obtained from the classical DCF method \citep{Crutcher2004}. First, we fit a Gaussian curve to the histogram of POL-2 polarization angles shown in the top panel of Figure~\ref{fig:fig5_scupol_histo} and find a dispersion $\delta \Phi_{obs} = 0.213$~radians ($12.2^{\circ}$). We then evaluate the dispersion $\delta \Phi_{err}$ due to instrumental errors using the mean uncertainty in polarization angle of $0.099$~radians ($5.7^{\circ}$) given in Section~\ref{sub:pol2_perseusb1}. This allows us to calculate the intrinsic angular dispersion $\delta \Phi = \sqrt{\delta \Phi_{obs}^2 - \delta \Phi_{err}^2} = 0.188$~radians ($10.8^{\circ}$). We then use Equation~\ref{eq:dcf}, assuming a correction factor $A=0.5$ \citep[e.g.,][]{Pattle2017,Soam2018,Kwon2018}, to derive a plane-of-sky magnetic field amplitude $B_{\text{pos}} \sim 230$~$\mu$G. This larger value for $B_{\text{pos}}$ suggests that a more appropriate correction factor for B1 would be $A \sim 0.25$. However, this derived field strength of $230$~$\mu$G could even be a lower limit (in the context of the classical DCF method) since the polarization vectors around B1-c are also included in the Gaussian fit, and so the appropriate correction factor to use would in fact be $A \lesssim 0.25$.

With the magnetic field amplitude $B_{\text{pos}}=120 \pm 60$~$\mu$G we have obtained from the angular dispersion analysis, it becomes possible to estimate the criticality criterion $\lambda_c$ of Perseus~B1 with Equation~\ref{eq:crit_ratio}. Using the hydrogen column density $N(\text{H}_2) = (4.7 \pm 0.9) \times 10^{22}$~cm$^{-2}$ derived previously, we find $\lambda_c = 3.0 \pm 1.5$. Since $\lambda_c > 1$, Perseus~B1 is a magnetically supercritical molecular cloud, i.e., magnetic pressure alone cannot support the cloud against gravity.

Perseus B1 is among a few molecular clouds with a detection of OH Zeeman splitting, and thus a measurement of its magnetic field's line-of-sight component. With observations of the OH lines at 1665~MHz and 1667~MHz using the Arecibo telescope and a beam width of $2.9$~arcmin, \citet{Goodman1989} found a line-of-sight amplitude of $27 \pm 4$~$\mu$G for the magnetic field towards IRAS 03301+3057 (B1-a). While this value might have been overestimated relative to the line-of-sight amplitude of the magnetic field at large scales \citep{Crutcher1993,Matthews2002}, it nonetheless supports the idea that the orientation of the magnetic field in B1 might be mostly parallel to the plane of the sky (i.e., an inclination $\theta < 15^{\circ}$ relative to the plane of the sky).

\begin{deluxetable*}{ccl}
\tablecaption{Derived magnetic and turbulent properties, and other related parameters in Perseus B1 \label{tab:houde_results}}
\tablewidth{0pt}
\tablecolumns{3}
\tablehead{\colhead{Parameter} & \colhead{Value} & \colhead{Description}}
\startdata
$\delta$ & $5.0 \pm 2.5$ arcsec & Turbulent correlation length scale\\
$N$ & $27.3 \pm 0.3$ & Number of beam-integrated turbulent cells along the line-of-sight\\
$\left\langle B_t^2 \right\rangle / \left\langle B_o^2 \right\rangle$ & $0.9 \pm 1.1$ & Turbulent-to-ordered magnetic energy ratio\\
$\left\langle B_t^2 \right\rangle / \left\langle B^2 \right\rangle$ & $0.5 \pm 0.3$ & Turbulent-to-total magnetic energy ratio\\
$a$ & $(2.4 \pm 0.2) \times 10^{-6}$~arcsec$^{-2}$ & First Taylor coefficient of the ordered auto-correlation function\\
$\delta V$ & $(2.9 \pm 1.1) \times 10^4$~cm~s$^{-1}$ & Velocity dispersion of the gas along the line-of-sight \tablenotemark{a}\\
$n(\text{H}_2)$ & $(1.5 \pm 0.3) \times 10^{5}$~cm$^{-3}$ & Mean number density of the gas \tablenotemark{a}\\
$N(\text{H}_2)$ & $(4.7 \pm 0.9) \times 10^{22}$~cm$^{-2}$ & Estimated column density for a cloud depth of $\sim 0.1$~pc\\
$\rho$ & $(7.0 \pm 1.4) \times 10^{-19}$~g~cm$^{-3}$ & Estimated density of the gas for a molecular weight $\mu=2.8$\\
$B_{\text{pos}}$ & $120 \pm 60$~$\mu$G & Plane-of-sky amplitude of the magnetic field\\
$\lambda_c$ & $3.0 \pm 1.5$ & Criticality ratio \tablenotemark{b}\\
\enddata
\tablenotetext{a}{ \citealt{Friesen2017}; GAS Consortium, in prep.}
\tablenotetext{b}{ \citealt{Crutcher2004}}
\end{deluxetable*}

\section{Discussion}
\label{sec:discussion}

\subsection{Morphology of the Magnetic Field}
\label{sub:morphology}

The magnetic field in Perseus B1, as shown in the right panel of Figure~\ref{fig:fig1_b1_polarization}, is seen to run roughly North-South (or $\sim 165^{\circ}$ East of North) across the whole region, including SMM3. The orientation of the vectors seen in Figure~\ref{fig:fig1_b1_polarization} (right) towards the bulk of the cloud (between B1-b N/S and SMM3) can be explained if B1 is part of a dense, slightly flattened cylindrical filament threaded perpendicularly by a large-scale magnetic field and viewed at an inclined angle to the line-of-sight \citep{Tomisaka2015}. While it may not be clear from Figure~\ref{fig:fig1_b1_polarization} alone, Perseus B1 is indeed part of a large filamentary structure extending towards the South-Western part of the map \citep{Chen2016}. Furthermore, magnetic field lines perpendicular to large-scale filaments have been hypothesized to funnel low density material into the striations (or sub-filaments) observed with \textit{Herschel} in and around molecular clouds \citep{Andre2014}. Alternatively, if the cloud is collapsing gravitationally, then the apparent curving of the field lines West of SMM3 could be the sign of an emergent hourglass morphology \citep[e.g.,][]{Girart2006}.

The largest discrepancy in the morphology of the large-scale magnetic field is seen towards the protostellar core B1-c, which is the source of a powerful molecular outflow viewed almost edge-on \citep{Matthews2006}. Indeed, the field turns more towards an East-West direction (or $\sim 120^{\circ}$ East of North) in the vicinity of B1-c, where it seems instead better aligned with the orientation of the protostellar outflow traced by the $^{12}$CO J=3-2 integrated intensity contour. In fact, the plane-of-sky component of the magnetic field towards B1-c is nearly parallel to the orientation of the outflow at $125^{\circ}$. In contrast, the local magnetic field direction is relatively well aligned with the mean field orientation in Perseus B1 ($\sim 165^{\circ}$) at the locations of the candidate first hydrostatic cores, and potentially less evolved, B1-bN ($\sim 155^{\circ}$) and B1-bS ($\sim 165^{\circ}$) objects \citep{Pezzuto2012,Gerin2017}, as well as at the previously identified young stellar objects associated with the submillimeter sources B1-a ($\sim 159^{\circ}$) and SMM3 ($\sim 158^{\circ}$), and to a lesser extent B1-d ($\sim 10^{\circ}$) and HH 789 ($\sim 180^{\circ}$) \citep{Bally2008}. This directional variation suggests that the magnetic field morphology is well ordered at large scales, but is potentially locally modified by the motion of the gas at smaller scales.

Perhaps the magnetic field orientation at B1-c originally followed the large-scale field of the molecular cloud, but was misaligned with the angular momentum of the initial prestellar core. As the core evolved, the magnetic field lines may have been ``dragged'' into a modified hourglass configuration \citep[e.g.,][]{Kataoka2012}. However, although hourglass structures have been seen toward some protostellar cores \citep[e.g.,][]{Girart2006, Hull2017b}, an alignment between magnetic field and outflow orientations does not appear to be a common occurrence \citep{Hull2014}.

Alternatively, the orientation of the magnetic field at B1-c could be explained by more complex field models which have been shown to produce comparable polarization patterns \citep{Franzmann2017}. Indeed, recent ALMA observations of the protostellar core Ser-emb~8 in Serpens Main suggest that the magnetic field of that object, which is similarly misaligned with the large-scale field of the rest of the molecular cloud in which it is embedded, may not possess an hourglass morphology at all \citep{Hull2017}. However, the protostellar core Serpens~SMM1 (also in Serpens Main) nevertheless shows evidence of having an hourglass field morphology while still being misaligned with the magnetic field at larger scales \citep{Hull2017b}. It would therefore be premature to assume that an observed misalignment in magnetic field orientations between core and cloud scales necessarily implies the absence of an hourglass field morphology.

Another peculiar property of B1-c is the orientation of the few polarization vectors found East from the protostellar core and along its outflow, as traced by the $^{12}$CO J=3-2 contour in Figure~\ref{fig:fig1_b1_polarization}. The inferred magnetic field orientation from the vectors found directly in the outflow's path ($\sim 160^{\circ}$) is in better agreement with the large-scale field in B1 ($\sim 165^{\circ}$) than with the field orientation towards B1-c itself ($\sim 120^{\circ}$). Magnetic field orientations that are nearly perpendicular to outflows at large scales are not expected from ideal hourglass field morphologies.

An alternative explanation would be that elongated dust grains found in the vicinity of the outflow are aligned mechanically by the flow of gas instead of radiatively. In this case, the polarization vectors would be parallel (and the inferred magnetic field orientation perpendicular) to the outflow orientation, regardless of the field morphology \citep{Gold1952a, Lazarian1997, Lazarian2007_review}, as is seen. This last scenario, however, has been shown to be unlikely even in the case of explosives outflows such as in Orion BN/KL \citep{Tang2010}.

Indeed, the original mechanical alignment proposed by \citet{Gold1952a} requires supersonic flows to be efficient, and it is particularly inefficient for suprathermally rotating grains (see \citealt{Lazarian1997}, \citealt{Das2016}). Thus, although its polarization pattern seems to be consistent with the observed polarization map, it is rather difficult to explain the high polarization degree ($\sim~15$ per cent) shown in Figure~\ref{fig:fig1_b1_polarization}. On the other hand, the MechAnical Torque (MAT) alignment mechanism proposed by \citet{Lazarian2007b} and numerically demonstrated by \citet{Hoang2018} predicts that the gas flow can efficiently align grains with the magnetic field. Specifically, the MAT mechanism predicts that the long-axis of the grains will be perpendicular either to the magnetic field or the gas flow. Therefore, the polarization vectors found along the outflow's lobes may reveal that the magnetic field in the flow is not much different from the large-scale magnetic field in the rest of the molecular cloud.

Finally, there is the possibility that we are mainly measuring the polarization from dust grains found in the cavity walls of the B1-c outflow. Indeed, it has been suggested that strong irradiation of outflow cavity walls can enhance the polarized emission of the associated dust grains through radiative torques \citep[e.g.,][]{Maury2018}. This scenario is supported by ALMA observations of B1-c (or Per-emb-29) \citep{Cox2018} which provide evidence for significantly improved grain alignment (with $P > 5$~per cent) along outflow cavities near the protostar. Although previous ALMA studies have shown that the magnetic field along comparable outflow cavities tend to be parallel to the outflow orientation \citep{Hull2017b,Maury2018,Cox2018} instead of perpendicular as observed eastward from B1-c in Figure~\ref{fig:fig1_b1_polarization}, their spatial resolutions were much smaller ($140$~au, $60$~au, and $100$~au respectively) than our resolution of $\sim 3500$~au. It could be that the dust grains with potentially enhanced polarized emission farther along the outflow cavity are instead tracing the large-scale field in the cloud, which would fit with the twisted field picture from \citet{Kataoka2012} where the polarization signature becomes less affected by the outflow the farther away you look from the central source.

\subsection{Magnetic and Turbulent Properties}
\label{sub:magnetic_turbulence}

In Section~\ref{sub:dispersion_results}, we derived the turbulent and magnetic properties of Perseus B1 from the angular dispersion analysis described by \citet{Houde2009} (see Figure~\ref{fig:fig7_dcf}). Specifically, we obtain a ratio of turbulent-to-total magnetic energy $\left\langle B_t^2 \right\rangle / \left\langle B^2 \right\rangle = 0.5 \pm 0.3$, which indicates that a large part of the magnetic energy in the cloud is found in the form of magnetized turbulence. This is larger than the ratio $\left\langle B_t^2 \right\rangle / \left\langle B^2 \right\rangle \sim 0.4$ found by \citet{Levrier2018} for the galactic magnetic field using Planck data. As a comparison, a previous study utilizing the angular dispersion analysis presented in Section~\ref{sub:dcf} found ratios of turbulent-to-total magnetic energy $\left\langle B_t^2 \right\rangle / \left\langle B^2 \right\rangle$ of, respectively, 0.6, 0.7 and 0.7 for the high mass star-forming regions W3(OH), W3 Main and DR21(OH) \citep{Houde2016}.

Since the ionized and neutral components of the gas in molecular clouds are typically well coupled, this magnetized turbulence is expected to be indistinguishable from the turbulence in the neutral gas as long as ambipolar diffusion remains negligible \citep[e.g.,][]{Krumholz2014}. Furthermore, the relatively large turbulent component of the magnetic field in B1 could be explained by the presence of at least five young stellar objects with confirmed molecular outflows (B1-a, B1-bS, B1-c, B1-d, and HH~789) in the main body of the cloud \citep{Hatchell2009}. Indeed, such outflows are among the most probable drivers of turbulence in molecular clouds \citep{Bally2008}. However, the signature of this protostellar feedback on the velocity dispersion of NH$_3$ does not appear to be as pronounced in B1 (GAS Consortium, in prep.) as it is in the more compact B59 in the Pipe nebula \citep[see Figure~9 in][]{Redaelli2017}, but a more detailed coherence analysis will be required to adequately investigate this effect.

The turbulent cells in B1 have a correlation length $\delta$ of $5.0 \pm 2.5$~arcsec, which for a distance of 295~pc represents a physical length of 1475~au. From Equation \ref{eq:turbulence}, we estimate that there are typically $\sim 30$ turbulent cells probed by the telescope's beam along the depth of the cloud ($0.1$~pc). The number of turbulent cells along the line-of-sight could potentially be greater in higher density regions, such as towards pre-stellar cores. This larger number would explain the observed depolarization effect seen in Figure~\ref{fig:fig2_depolarization} (top) as the Stokes $I$ intensity increases, which can be roughly understood as an increase in the dust column density.  Indeed, an increased number of turbulent cells is expected to randomize dust orientations along the line-of-sight, and thus decrease the measured fraction of polarization $P$. Additionally, and perhaps counter-intuitively, numerical simulations by \citet{Cho2016} have also shown that the averaging of a high number of turbulent cells along the line-of-sight could preserve the appearance of a well-ordered field morphology at large scales, which is an effect initially proposed by \citet{Jones1992}.

In Section~\ref{sub:dispersion_results}, we also find a plane-of-sky amplitude $B_{\text{pos}} = 120 \pm 60$~$\mu$G for the magnetic field, and a criticality criterion $\lambda_c = 3.0 \pm 1.5$. Although this magnetic field amplitude is relatively weak when compared to the fields found in high mass star-forming regions such as Orion~A (where $B_{\text{pos}} \gtrsim 1.0$~mG) \citep[e.g.,][]{Houde2009, Pattle2017} or in hub-filament structures such as IC~5146 (with $B_{\text{pos}} \sim 0.5$~mG) \citep[e.g.,][]{Wang2018arxiv}, it is either comparable to or larger than the field strengths ($B_{\text{pos}} \lesssim 100$~$\mu$G) typically found in low-mass prestellar cores \citep[e.g.,][]{Crutcher2004, Kirk2006, Liu2019arxiv}. Above all, these results indicate that Perseus B1 is a supercritical molecular cloud (i.e., magnetic pressure alone cannot support the cloud against gravity). The criticality criterion $\lambda_c$ defined by Equation~\ref{eq:crit_ratio}, however, may be overestimated due to geometric effects. Indeed, \citet{Crutcher2004} find that, on average, the effective criticality criterion is $\overline{\lambda_{c}} \approx \lambda_c / 3$. In the case of B1, this adjustment would lead to $\overline{\lambda_{c}} \approx 1.0$, which is the theoretical limit at which the cloud would be subcritical. 

Since the inclination of the magnetic field in B1 can be calculated using published Zeeman line splitting measurements (see Section~\ref{sub:dispersion_results}), we can better estimate the effect of geometry on the criticality criterion $\overline{\lambda_c}$. Assuming that the line-of-sight component obtained by \citet{Goodman1989} ($27 \pm 4$~$\mu$G) is not an overestimation at large scales, we find an inclination $\theta = 12^{\circ}$ relative to the plane of the sky and an amplitude $B_{\text{tot}} \approx 125$~$\mu$G for the total magnetic field when combined with the plane-of-sky amplitude $B_{\text{pos}} = 120 \pm 60$~$\mu$G found in Section~\ref{sub:dispersion_results}. If the cloud can also be approximated as a mostly prolate filament with a cylindrical symmetry, which is a reasonable assumption for a relatively weak magnetic field in a dense filament, then we get $\overline{\lambda_c} \approx \lambda_c$. We therefore find it likely that Perseus~B1 is indeed supercritical by a factor $\sim 3$, although we cannot rule out if a combination of magnetic pressure and turbulence would be sufficient to significantly slow down the fall of additional material onto the central clump.  

\subsection{Polarization Fraction and Grain Alignment}
\label{sub:depolarisaton}

Fundamentally, the fraction $P$ of polarization can be understood as the alignment efficiency of a mixture of dust grains in the interstellar medium. Even though this fraction $P$ can be affected by purely environmental factors such as the number of integrated turbulent cells along the line-of-sight and complex magnetic field geometries, or even instrumental factors such as molecular contamination (see Appendix~\ref{sub:contamination}), it is intrinsically linked to the models of grain alignment. 

Specifically, the contribution to the continuum emission of different grain sizes and compositions in the dust mixture could explain the apparent dependence of $P$ on the wavelength at far-infrared and submillimeter wavelengths \citep{Vaillancourt2012}. For example, grain growth in cold high density regions may lead to very large dust grains, with sizes $a \gtrsim 1.0$~$\mu$m \citep[e.g.,][]{Pagani2010}, which align less efficiently through radiative torques than the typical grains ($a \sim 0.1$~$\mu$m) found in molecular clouds \citep{Hoang2009}. This scenario could potentially explain the apparent drop in polarization fraction~$P$ seen in Figure~\ref{fig:fig3_depolarization} above a visual extinction $A_V > 200$~mag, as well as towards B1-c, since there is significant evidence for grain growth across Perseus~B1 \citep{Sadavoy2013, Chen2016}.

Furthermore, since the RAT theory of grain alignment depends on the stellar radiation field incident on the grains, the alignment efficiency is expected to be smaller towards regions with high dust opacities (e.g., dense prestellar cores) \citep{Andersson2015}. This effect would potentially explain the apparent minimum $P$ of $\sim$1~per cent seen both in Figure~\ref{fig:fig2_depolarization} (top) and by \citet{Matthews2002} for the highest opacity regions of the cloud, which in the case of Perseus B1 are associated with embedded young stellar objects such as the first hydrostatic core candidates B1-b N/S (see Figure~\ref{fig:fig1_b1_polarization}). This alignment efficiency, however, is expected to improve again if there is a significant source of radiation, such as a protostar, within the core itself. Such a scenario would explain the shallower than expected power index $\beta \sim -0.5$ given in Section~\ref{sub:pol2_perseusb1} for the relation between the polarization fraction~$P$ and the visual extinction~$A_V$ in Perseus~B1.

Nevertheless, B1-c, which is known to be a bright and warm protostellar core \citep{Sadavoy2013}, also has among the lowest polarization fractions measured by POL-2 for B1. This behavior suggests that we may not be resolving the improved grain alignment efficiency seen by ALMA near the protostar \citep{Cox2018}. Indeed, \citet{Jones2016} previously observed such an effect when comparing single-dish and interferometric polarization data of the protostellar core G034.43+00.24 MM1. Alternatively, it could be that factors other than alignment efficiency need to be taken into account to explain the polarization towards this object. 

As an example, previous studies have found an inverse correlation between the polarization fraction~$P$ and the local dispersion of magnetic field orientations at several scales in molecular clouds \citep{Planck2015XIX, Planck2015XX, Fissel2016, Koch2018}. Such a measure towards B1-c would support the hypothesis of a complex but unresolved polarization structure, and higher resolution observations using interferometric facilities would provide further evidence to confirm or infirm this scenario. However, while there exist ALMA data of the linear polarization towards B1-c, only the most highly polarized emission is likely to have been recovered due to the short integration time ($8$~minutes) of these observations \citep{Cox2018}. A deeper ALMA polarization map of B1-c might therefore reveal a more complex magnetic field structure comparable to those observed in similar protostellar cores  \citep[e.g.,][]{Hull2017,Hull2017b,Maury2018}.

\section{Conclusion}
\label{sec:conclusion}

We have observed the 850~$\mu$m linear polarization towards the B1 clump in the Perseus molecular cloud complex using the POL-2 polarimeter as part of the BISTRO survey at the JCMT. We have also compared the resulting polarization map with previously published SCUPOL observations of B1 from \citet{Matthews2009} to illustrate the improvements brought by the increased sensitivity and reliability of POL-2 over its predecessor. From the POL-2 observations, we have inferred the plane-of-sky morphology of the magnetic field in Perseus B1 by rotating the 850~$\mu$m polarization vectors by 90$^\circ$ assuming the dust grains are aligned by radiative torques \citep[e.g.,][]{Andersson2015}. The plane-of-sky component of the magnetic field in most of the cloud is orientated in a North-South direction (or $\sim 165^{\circ}$ East of North), except towards the protostellar core B1-c where it turns more East-West in better agreement with the orientation of its associated molecular outflow. 

We have also plotted the polarization fraction $P$ and the de-biased polarized intensity $I_P$ as a function of the Stokes $I$ total intensity. Specifically, we have fitted a power-law to the relationship between $P$ and $I$, and we find a power index $\alpha \sim -0.9$ in agreement with other BISTRO studies. There exists a clear trend in Perseus B1 of decreasing polarization fraction $P$ as a function of increasing Stokes $I$, although the polarized intensity $I_P$ itself appears to increase steadily. Such a behavior is likely linked to depolarization effects towards higher density regions, such as a complex field geometry, a low efficiency of grain alignment, or an increased number of turbulent cells along the line-of-sight. 

Similarly, we have plotted the polarization fraction $P$ as a function of the visual extinction~$A_V$ in Perseus~B1, and fitted a power-law between the two parameters. We find a power index $\beta \sim -0.5$, which is a shallower value than those previously found in starless cores with comparable extinction measurements ($A_V>20$). This shallow power index $\beta \sim -0.5$ could therefore be explained by improved grain alignment due to the radiation from embedded young stellar objects in the cloud.

We have applied the angular dispersion analysis developed by \citet{Houde2009} to the POL-2 850~$\mu$m  polarization map of Perseus B1. By fitting the angular dispersion function, we have measured a turbulent magnetic correlation length $\delta$ of $5.0 \pm 2.5$~arcsec, which for a distance of 295~pc represents a physical length of $\sim 1500$~au, and a turbulent-to-total magnetic energy ratio of $0.5 \pm 0.3$ inside the cloud. Such a large ratio indicates that a significant part, if not most, of the magnetic energy in the cloud is found in the form of magnetized turbulence. Additionally, using an effective cloud depth of $\sim 0.1$~pc, we have evaluated that there are typically $\sim 30$ beam-integrated turbulent cells along the line-of-sight across B1.   

With an updated version of the Davis-Chandrasekhar-Fermi method, we have evaluated the plane-of-sky amplitude of the magnetic field in Perseus B1 to be $B_{\text{pos}} = 120 \pm 60$~$\mu$G. From this amplitude, we have estimated the magnetic criticality criterion in this cloud to be $\lambda_c = 3.0 \pm 1.5$. We also found with measurements of OH Zeeman line splitting that the orientation of the magnetic field is nearly parallel to the plane of the sky, and thus this criticality criterion is unlikely to be overestimated due to geometric effects. Perseus~B1 is therefore a magnetically supercritical molecular cloud. 

Finally, our findings show that the angular dispersion analysis presented by \citet{Houde2009} can be successfully applied to POL-2 observations of nearby star-forming regions. It will therefore be possible in future works to expand this analysis to a representative sample of molecular clouds in order to systematically quantify, and compare, their magnetic and turbulent properties. This illustrates how the BISTRO survey has the potential to provide us with unparalleled insight into the roles of magnetic fields and turbulence in the physical processes leading to the formation of stars and their planets.

\section*{Acknowledgements}

We would like to thank the staff of the East Asian Observatory for their invaluable support in the completion of the BISTRO survey. We also wish to thank the people of Hawai'i for granting us access to the unique geographical site of the Maunakea observatory. Furthermore, we are grateful to the GAS Consortium for generously granting us access to their spectroscopic data. We also thank the anonymous reviewer for their helpful and detailed comments. Finally, we thank B.-G. Andersson, Kelvin Au, Jordan Guerra Aguilera, James Lane, Anna Ordog, Am\'elie Simon, Ian Stephens, and Julien Vandeportal for helpful discussions.

This research was conducted in part at the SOFIA Science Center, which is operated by the Universities Space Research Association under contract NNA17BF53C with the National Aeronautics and Space Administration.

The James Clerk Maxwell Telescope is operated by the East Asian Observatory on behalf of The National Astronomical Observatory of Japan; Academia Sinica Institute of Astronomy and Astrophysics; the Korea Astronomy and Space Science Institute; Center for Astronomical Mega-Science (as well as the National Key R\&D Program of China with No. 2017YFA0402700). Additional funding support is provided by the Science and Technology Facilities Council of the United Kingdom and participating universities in the United Kingdom and Canada.

SCUBA-2 and POL-2 were built through grants from the Canada Foundation for Innovation. This research used the facilities of the Canadian Astronomy Data Centre operated by the National Research Council of Canada with the support of the Canadian Space Agency. This research has also made use of the \textsc{simbad} database and of NASA's Astrophysics Data System Bibliographic Services. The Starlink software \citep{Currie2014} is currently supported by the East Asian Observatory.

Miju Kang was supported by Basic Science Research Program through the National Research Foundation of Korea (NRF) funded by the Ministry of Science, ICT \& Future Planning (NRF-2015R1C1A1A01052160). Woojin Kwon was supported by Basic Science Research Program through the National Research Foundation of Korea (NRF-2016R1C1B2013642). C.W.L. was supported by the Basic Science Research Program through the National Research Foundation of Korea (NRF) funded by the Ministry of Education, Science and Technology (NRF-2016R1A2B4012593). Keping Qiu is supported by National Key R\&D Program of China No. 2017YFA0402600, and acknowledges the support from National Natural Science Foundation of China (NSFC) through grants U1731237, 11473011, 11629302, and 11590781.

\software{\textsc{Starlink} \citep{Currie2014, POL2_Cookbook, Chapin2013}, The IDL Astronomy User's Library \citep{Landsman1993}.}

\bibliographystyle{aasjournal}
\bibliography{b1_polarization}

\appendix

\section{Effect of Molecular Contamination}
\label{sub:contamination}

Another effect which may influence the measured fractions of polarization is the contribution from molecular line emission at submillimeter wavelengths. The $^{12}$CO J=3-2 molecular line in particular has been shown in some special cases to be a significant source of contamination in SCUBA-2 continuum observations at 850~$\mu$m \citep{Drabek2012}. While relatively rare, high levels of $^{12}$CO J=3-2 line contamination ($>$10~per cent) in star-forming regions are usually associated with molecular outflows from young stellar objects \citep[e.g.,][]{Chen2016,Coude2016}. This behavior occurs in SCUBA-2 observations of Perseus B1, where \citet{Sadavoy2013} found $^{12}$CO J=3-2 line contamination levels of 90~per cent in the outflows of B1-c, 15~per cent in the central region of B1, and $<~1$~per cent in the rest of the cloud. 

It is important to note that HARP, SCUBA-2, and POL-2 are not sensitive to the same spatial scales due to their different observing strategies. Specifically, SCUBA-2 observations for the JCMT Gould Belt Survey were taken using a PONG 1800 observing mode that is sensitive to larger spatial scales than the Daisy mode used for POL-2 \citep{Chapin2013,Friberg2016}. We therefore expect contamination levels for POL-2 to be different than those previously measured for SCUBA-2 alone, but nonetheless still confined to molecular outflows if present. Similarly, HARP observations are sensitive to larger angular scales than those from SCUBA-2, and they had to be spatially filtered during data reduction to be subtracted accurately from the 850~$\mu$m maps of the JCMT Gould Belt Survey \citep[e.g.,][]{Mairs2016}. Such a subtraction procedure for $^{12}$CO J=3-2 molecular line contamination could potentially be adapted for future analyses of BISTRO observations. 

The emission from the $^{12}$CO J=3-2 molecular line can be weakly linearly polarized by magnetic fields through the Goldreich-Kylafis effect \citep{Goldreich1981,Goldreich1982}. Observational evidence, however, suggests that this polarization is only on the order of 1~per cent for single-dish observatories \citep[e.g.,][]{Greaves1999, Forbrich2008}. Such a level of polarization would only be detectable by POL-2 in extreme cases of molecular contamination, such as the unlikely scenario of a $\sim 1.3$~Jy beam$^{-1}$ submillimeter source with a $^{12}$CO J=3-2 contamination level of 90~per cent (assuming a $3 \, \sigma$ detection threshold of $I_P \sim 12$~mJy beam$^{-1}$, and the maximum contamination fraction measured by \citealt{Sadavoy2013}). If there is significant contamination from the $^{12}$CO J=3-2 molecular line in POL-2 observations at 850~$\mu$m, it is reasonable to assume that this additional contribution to the continuum flux is unpolarised. Therefore, the effect of contamination will be to overestimate the Stokes $I$ total intensity while the Stokes $Q$ and $U$ parameters remain unchanged.  

In other words, molecular contamination from the $^{12}$CO J=3-2 molecular line will lead to an underestimation of the polarization fraction $P$, but the polarization angle $\Phi$ will be unaffected if the instrumental polarization is properly taken into consideration. This effect is thus unlikely to influence our characterization of the magnetic and turbulent properties of Perseus~B1, although it could potentially affect the polarization fraction $P$ plotted in Figure~\ref{fig:fig2_depolarization} (top). Such possible contamination may need to be taken into account for future, more detailed analysis of grain alignment efficiency using POL-2 data. 

Finally, it is important to note that the Goldreich-Kylafis effect might nonetheless be important for polarimetric observations using interferometers such as the SMA. Indeed, \citet{Ching2016} measured polarization fractions up to 20~$\%$ for the $^{12}$CO J=3-2 emission towards the IRAS~4A protostellar outflow. In such cases, continuum measurements of the Stokes~$Q$ and $U$ parameters are likely to be affected by strong $^{12}$CO line contamination of the Stokes~$I$ total intensity.


\end{document}